\documentclass[preprint,sort&compress,authoryear]{elsarticle}

\usepackage{amsmath,amsthm,amssymb,latexsym,amscd}
\usepackage{subfigure} 
\usepackage{booktabs}
\usepackage[list-final-separator={, and }]{siunitx}
\usepackage[shadow,textsize=small,textwidth=2.5cm]{todonotes}
\usepackage{stmaryrd}
\newtheorem{remark}{Remark}
\usepackage{geometry}
\usepackage{mathtools}
\usepackage{algorithm}
\usepackage{algpseudocode}
\algtext*{EndIf}
\algtext*{EndWhile}
\algtext*{EndFor}

\usepackage{color}
\usepackage[title]{appendix}
\usepackage{nomencl}
\usepackage{bm}
\usepackage{lineno}
\usepackage{cancel}

\usepackage{amsmath,amsthm,amssymb,latexsym,amscd}
\usepackage{lineno}
\usepackage[defaultcolor=red]{changes}

\usepackage{todonotes}

\newcommand{\mrm}{\mathrm}
\newcommand{\Gc}{G_\mrm{c}}

\newcommand{\eps}{\varepsilon}

\newcommand{\mathd}{\mathrm{d}}
\newcommand{\mbf}[1]{{\mathbf{#1}}}

\newcommand{\mbfs}{\boldsymbol}
\newcommand{\Tr}[1]{\text{Tr} \left( #1\right) }

\usepackage{hyperref}
\hypersetup{
	colorlinks=true,
	linkcolor=blue,
	filecolor=magenta,      
	urlcolor=cyan,
}

\begin{document}
	\begin{frontmatter}
        \title{A dual--continuum phase-field model for hydraulic fracturing: Viscosity-dominated regime and fluid lag}
		\author[MUL]{Tao You}
		\ead{tao.you@unileoben.ac.at}
		\author[MUL]{Keita Yoshioka}
		\ead{keita.yoshioka@unileoben.ac.at}
		\address[MUL]{Department Geoenergy, Technical University of Leoben, Leoben 8700, Austria}
		
\begin{abstract}
The phase-field model regularizes sharp fractures into a diffuse representation, blurring the boundary between the fracture and the intact material. This blurring makes it difficult to capture distinct domain processes in hydraulic fracturing, where Reynolds flow governs the fracture and Darcy flow describes the surrounding porous matrix. Consequently, the blurred delineation artificially smears the pressure field across the fracture--matrix interface, which is acceptable in toughness-dominated hydraulic fracturing regimes where pressure drops within the fracture are negligible. However, in viscosity-dominated regimes, typically for actual subsurface injections due to high injection rates, the fluid pressure drops more drastically, and the fluid front may even lag behind the propagating fracture tip, a phenomenon that a smeared pressure field cannot capture. Despite its relevance, the viscosity-dominated regime has not been addressed by any existing phase-field models to date, likely due to its numerical instability. In this study, we propose a dual--continuum phase-field model based on double-porosity microporomechanics that explicitly separates mesoscale crack pressure from micropore pressure. The framework provides a variationally consistent formulation alongside phase-field--dependent poroelasticity. To ensure the numerical stability of the hydromechanical coupling, a fixed-stress split scheme is modified for two independent fluid pressures, while a variational inequality constraint is applied to reproduce fluid lag. Verified against the closed-form solutions in toughness-dominated, viscosity-dominated, and early-time transitional regimes, the model accurately captures complex fluid flow behavior and transient fluid lag within the fracture, and opens a new frontier for applying phase-field models to realistic viscosity-dominated hydraulic fracturing.

		\end{abstract}
		
	\end{frontmatter}
	
	
\section{Introduction}

\label{sec:introduction}

\begin{figure}
    \centering
    \includegraphics[width=0.8\linewidth]{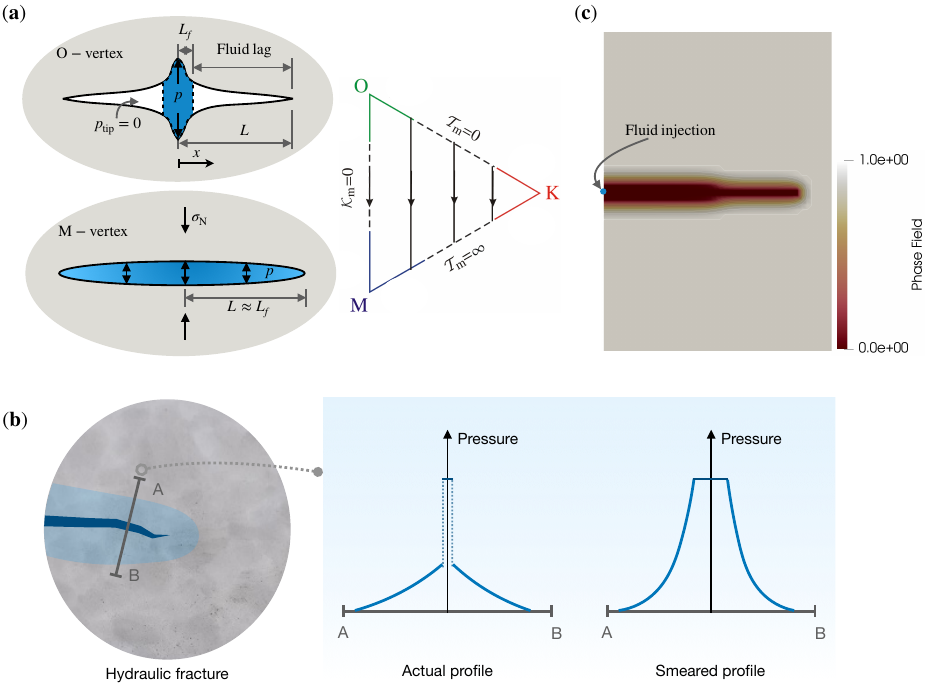}
    \caption{ (a) Hydraulic fracture propagation regimes, adapted from~\cite{garagash2006propagation}, where the parametric space of the three vertices of the limiting regimes: toughness-dominated (K), viscosity-dominated (M), and early-time (O). $p_\mathrm{tip}$ is the pressure in the fluid lag zone, and $\sigma_\mathrm{N}$ is the far-field confining pressure. (b) Pressure profiles across the hydraulic fracture. An actual pressure profile and a numerical smeared profile from a phase-field model. (c) The phase-field profile instability (broadening) driven by the high fluid pressure across the diffuse pressure profile and a negligibly small fracture toughness. }
    \label{fig:parametric_space}
\end{figure}

Fracturing in porous media is central to numerous engineering applications, such as hydrogel failure~\citep{baumberger2006solvent, bouklas2015effect}, biological tissue rupture~\citep{dougan2022cavitation}, and both natural~\citep{mori2022three} and anthropogenic~\citep{Legarth2005,king2010thirty,fu2017influence,plua2024situ} hydraulic fracturing of geological formations.
The behavior of a hydraulic fracture is governed by the interplay between solid skeleton deformation, multi-scale (fracture and pore--space) fluid flow, and crack propagation. 
Beyond laboratory~\citep{bunger2008experimental,bunger2013comparison,Ha2018,de2017laboratory} or field~\citep{Warpinski1987,Legarth2005,Economides2000} investigations, dimensionless scaling enables a systematic analysis of fluid-driven fracture mechanics~\citep{nilson1981gas, detournay2003near}.
\cite{garagash2006propagation} identified three different regimes---the toughness--dominated, the viscosity--dominated, and the early--time regimes---based on a dimensionless toughness $\mathcal{K}_\mathrm{m}$ and a dimensionless time $\mathcal{T}_\mathrm{m}$ (Fig.~\ref{fig:parametric_space}a).
The toughness-dominated regime ($\mathrm{K}$-vertex) is characterized by the dominant role of fracture toughness, while the viscosity-dominated regime ($\mathrm{M}$-vertex) is characterized by the dominant role of fluid viscosity, high far-field stress $\sigma_\mathrm{N}$, and prolonged fluid injection~\citep{lecampion2007implicit}. 
The early-time regime ($\mathrm{O}$-vertex) corresponds to the solution for small confining pressure and low toughness, in which a pronounced fluid lag zone develops near the fracture tip (Fig.~\ref{fig:parametric_space}a). 
This lag zone is either filled by fluid vapor (in impermeable medium) or by pore fluid drawn from the surrounding matrix (in permeable medium).

Despite recent advances, simulating hydraulic fractures remains challenging due to the inherent complexities of multiphysics coupling, moving fracture paths, and transient fluid fronts~\citep{lecampion2007implicit,lecampion2018numerical}. 
To address complex fracture paths, the variational phase-field model, regularization of the variational approach to brittle fracture~\citep{Francfort1998} through a phase-field variable~\citep{Bourdin2000}, offers distinct advantages, as it can capture crack branching, merging, and nucleation in a unified variational setting without explicit surface tracking. 
Subsequent algorithmic developments~\citep{bourdin2007numerical,miehe2010phase,Tanne2018} have demonstrated the robustness and versatility of this framework.
The approach has since been extended beyond brittle fracture to cohesive fracture~\citep{Verhoosel2013,wu2018length}, dynamic fracture~\citep{Borden2012phase,Schluter2014}, thermal cracking~\citep{Bourdin2014}, fatigue fracture~\citep{Carrara2020,sarem2025micromechanics}, and hydraulic fracturing -- the main focus in this study.
\cite{bourdin2012variational} first extended the variational fracture framework to hydraulic fracturing, and subsequent contributions by~\cite{Wheeler2014,Mikelic2015_CompGeo,Mikelic2015_Multi,Lee2016_adaptive,Lee2017_ls,wheeler2020ipacs,miehe2015minimization,miehe2016phase,Wilson2016,Yoshioka2016} established diverse numerical methods and convergence analyses for phase-field hydraulic fracture in poroelastic media. 
\cite{you2023onporo} recently proposed a micromechanics-based approach to derive effective, damage-dependent poroelastic properties for fractured porous media within a single-continuum framework. 
They demonstrated that the proposed model reproduces crack opening displacements more accurately than existing phase-field hydraulic fracture models, irrespective of the pressure profiles or the values of the Biot coefficient.
Beyond a single-fluid quasi-static hydraulic fracturing, further contributions include dynamic hydraulic fracturing~\citep{Ehlers2017,feng2021phase}, heterogeneous and layered formations~\citep{Xia2017,zhuang2020hydraulic,you2025remote}, and multi-fluid fracturing~\citep{lee2018phase,liu2025variational,li2025unified}.

In phase-field models, a diffuse fracture representation blurs the interface between the fracture and the porous media. 
Resolving this interface is critical in hydraulic fracturing because the domains are governed by different physical laws.
Fluid flow within the fracture is typically idealized as flow between parallel plates, where the flow rate is proportional to the cube of the crack opening (a relation known as Reynolds flow). 
Meanwhile, fluid flow in the surrounding porous medium is normally described by Darcy flow, where the flow rate is proportional to the matrix permeability.
To apply these two distinct hydraulic flows to each of the domains, various approaches have been proposed. 
One of the most popular, and intuitive, approaches is to use the phase-field value to delineate the domains. 
For instance, \cite{Mikelic2015_Multi, Mikelic2015_NonLin} used the phase-field variable to transition between the two flow regimes, while \cite{Yoshioka2016} applied two distinct permeabilities for the matrix and fracture based on the phase-field value.
Along a similar line, a level-set function based on the phase-field variable have been adopted to average the two flow behaviors by many~\citep{Lee2017_ls, li2019numerical,wheeler2020ipacs, liu2020investigation, li2021phase, xu2022phase,zhuang2022three, SHAHOVEISI2024117113}.
\cite{santillan2017phase} separated the fracture and reservoir domains based on the phase-field value, where the fracture domain is explicitly extracted and discretized using one-dimensional line elements to form a hybrid system with the surrounding two-dimensional reservoir elements. 
In their approach, however, this one-dimensional fracture domain must be updated dynamically as the phase-field profile evolves.
Another hybrid method was proposed by \citet{costa2022multi}, who used the phase-field model to capture fracture propagation; once a crack fully develops, the fractured domain transitions to a discrete contact formulation.
Alternatively, \citet{Wilson2016} used the phase-field variable to scale fluid viscosity and recover Reynolds flow within the fracture, whereas \citet{Chukwudozie2019} leveraged phase-field calculus to homogenize the distinct domains. 
By contrast, the frameworks in \citet{Heider2017, Ehlers2017} rely on the theory of porous media to govern the transition between Darcy flow in the reservoir and Navier-Stokes flow in the fracture.
Furthermore, despite its simple implementation, modifying the permeability as a function of the phase-field and local deformation has proven to be an effective way to capture enhanced flow within fractures~\citep{miehe2015minimization, you2023onporo}.

When considering these various treatments, we must keep two factors in mind. 
The first is the physical dimensions of hydraulic fractures; while fracture length typically spans tens of meters, the aperture is merely 0.001 to 0.01~m\footnote{For example, using the closed-form expression of internally pressurized crack opening of~\cite{Sneddon-Lowengrub-1969a}, the maximum crack opening displacement is given as $2pl/E$ with $p$, $l$, and $E$ being the internal fluid pressure, fracture half length, and Young's modulus. Using typical values of $p$=1~MPa, $l$=10~m, and $E$=10~GPa, the maximum aperture is 0.002~m.}. 
Even if a fracture is discretized with an element size of 0.1~m in numerical simulations, which may already be too computationally expensive to resolve a fracture spanning tens of meters, the physical fracture opening (aperture) remains far smaller than the element size. 
Because the phase-field transition zone spans 5 to 10 elements (depending on the characteristic length and the specific phase-field model applied), delineating the fracture based on the phase--field variable or a level--set function would yield an ``aperture'' that is enormous compared to its physical size.
Consequently, this nonphysical void space complicates the fluid mass balance.
The second factor is the fluid pressure profile.
Despite different approaches to the transition between the two distinct flow regimes (porous medium and fracture), all existing models treat pressure as a continuous field across the fracture-matrix interface\footnote{An exception is the work of \citet{santillan2017phase}; however, their approach sacrifices the primary advantages of the phase-field method, as the explicit fracture elements must be updated adaptively.}. 
Consequently, the pressure field becomes smeared (Fig.~\ref{fig:parametric_space}b).
This smeared pressure field appears to yield reasonable fluid pressure responses at least in the toughness--dominated hydraulic fracturing where the pressure drop within the fracture is negligible~\citep{chukwudozie2013variational}.
Conversely, under a high injection rate, high fluid viscosity, or low fracture toughness, hydraulic fracturing transitions into the viscousity-dominated regime. 
In this regime, fracture toughness is negligibly small compared to the viscous dissipation. 
Given a negligibly small fracture toughness combined with a smeared continuous fluid pressure, local driving forces around the fluid injection becomes excessive.
Consequently, damage tends to evolve in the transverse direction rather than out of  the fracture tip (Fig.~\ref{fig:parametric_space}c), causing numerical instability.
Although this type of instability has been systematically studied in simulations of moving phase interface (e.g., \cite{zhang2023phase,feyen2023quantitative}), it has not yet been addressed in variational phase-field models for fracture.
Furthermore, dominant viscous dissipation causes a drastic pressure drop within the fracture (as shown as the actual profile in Fig.~\ref{fig:parametric_space}b), and the fluid front may even lag behind the propagating fracture tip (Fig.~\ref{fig:parametric_space}a). 
This distinct pressure behavior and fluid lag cannot be fundamentally captured by a smeared pressure representation.

Because of the high injection rates typically applied in subsurface operations, hydraulic fracturing field applications are predominantly governed by the viscous-dominated regime.
Yet, thus far, no phase-field model has fully addressed the viscous-dominated regime (M-vertex).
\cite{Wilson2016} investigated a regime close to the M-vertex with $\mathcal{K}_\mathrm{m} = 0.97$ and \cite{santillan2017phase} investigated a state with $\mathcal{K}_\mathrm{m} = 1.43$.
Despite their claims, however, neither of the $\mathcal{K}_\mathrm{m}$ values in these studies was sufficiently small to capture the M-vertex regime ($\mathcal{K}_\mathrm{m} < 0.7$)~\citep{garagash2005plane}, and the fluid lag was not resolved in their works. 


To address the existing phase-field model's limitations in the viscous-dominated regime, this study introduces a dual--continuum phase-field fracture model based on two-scale microporomechanics (also known as double-porosity media), which explicitly distinguishes the pressure within the mesoscale cracks from the pressure in the micropores. 
Natural rock formations commonly exhibit porosity at two distinct scales: a network of mesoscale cracks that constitute the primary conduits for rapid fluid flow, and a micropore space within the intact matrix that governs fluid storage and diffusive exchange between the two pore systems.
The double-porosity concept was introduced by~\cite{barenblatt1960basic} for fissured porous media and formalized as a dual-continuum flow model by~\cite{warren1963behavior}.
Subsequent work incorporated the coupling between the two pore pressures and the solid deformation~\citep{coussy2004poromechanics,Biot1941}, and the double-porosity model has found widespread application in reservoir geomechanics~\citep{borja2009effective,choo2016hydromechanical}.
Despite this extensive body of literature, a dual-continnum framework has not been extended to phase-field models for propagating fractures.
In this paper, we propose a  phase-field model for hydraulic fracturing in a two-scale double-porosity medium, building on the microporomechanics of~\cite{pichler2010cracking,dormieux2006microporomechanics}.
The principal contributions of this study are: 
\begin{enumerate}[(i)]
    \item a variationally consistent phase-field model for hydraulic fracturing in two-scale double-porosity media in which the effective elastic tensor, the Biot coefficient tensors, and the Biot moduli are derived rationally from microporomechanics;
    \item the relaxation of pressure continuity assumption across the fracture using two independent pressure variables ($p_\mathrm{c} \neq p_\mathrm{p}$), thereby preventing artificially high local driving forces;
    \item a variational inequality formulation that enforces non-negative fluid pressures in the fracture, providing a flexible treatment of the fluid lag without explicit tracking of the fluid front;
    \item a fixed-stress split scheme for the two-pressure hydromechanical system that accounts for the coupling between the two pore networks; and
    \item successful numerical verification against analytical solutions in the toughness-dominated (K-vertex), viscosity-dominated (M-vertex), and early-time transitional regimes (O-vertex).
\end{enumerate}

The remainder of this paper is organized as follows.
Section~\ref{model} presents the theoretical framework, including the microporomechanics-based state equations and the phase-field fracture formulation.
Section~\ref{sec:solution} details the solution schemes, comprising the fixed-stress split and the variational inequality formulation for the fluid lag.
Section~\ref{sec:examples} presents the numerical verification studies and double-porosity simulations.
Concluding remarks are offered in Section~\ref{sec:conclusion}. 

Throughout the paper, the following notations are used.
Tensorial product of any second--order tensors $\boldsymbol{A}$ and $\boldsymbol{B}$ and fourth--order tensor $\mathbb{C}$ are $\boldsymbol{A}:\boldsymbol{B}=A_{ij}B_{ij}$ and $\mathbb{C}:\boldsymbol{B}=\mathbb{C}_{ijkl}\boldsymbol{B}_{kl}$. 
The symbol $\left\Vert \boldsymbol{A}\right\Vert=\sqrt{\boldsymbol{A}:\boldsymbol{A}}$ is used to calculate the norm of any second--order tensor $\boldsymbol{A}$. 
The identity tensors for the second and fourth order are denoted by $\mbfs{\delta}$ and $\mathbb{I}$, respectively.
The fourth order projector $\mathbb{J}$ and $\mathbb{K}$ are expressed in the component form as $J_{ijkl}=\delta _{ij}\delta _{kl}/3$ and $K_{ijkl}=(\delta _{ik}\delta_{jl}+\delta _{jl}\delta _{jk})/2-\delta _{ij}\delta _{kl}/3$, respectively. 
The trace operator $\Tr{\cdot}$ is defined as $\Tr{\mbfs{A}} = \mbfs{\delta}: \mbfs{A}$.
$\nabla (\cdot )$ is the gradient of $(\cdot )$, $(\cdot )^\mrm{T}$ is the transpose of $(\cdot )$, and $\nabla^\mrm{s} (\cdot ) $ is the symmetric part of the gradient of $(\cdot )$ defined as $\nabla^\mrm{s} (\cdot ) := \frac12 (\nabla (\cdot ) + \nabla^\mrm{T} (\cdot ))$. 
The overhead dot symbol $\dot{A}$ indicates the time differentiation of variable $A$.

\section{Theoretical frameworks} \label{model}

\subsection{State equations}
\begin{figure}[htp!]
		\centering
		\includegraphics[width=0.8\linewidth]{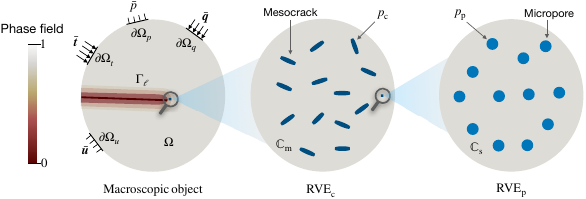}
		\caption{Schematic illustration of the scale separation of the cracked porous media where the meso-cracks and micro-pores are filled with fluid.
        \textit{Left}: macroscopic domain $\Omega$ containing a phase-field regularized macrocrack $\Gamma_\ell$. The boundary is partitioned into complementary pairs: prescribed traction $\bar{\boldsymbol{t}}$ on $\partial\Omega_t$ and prescribed displacement $\bar{\boldsymbol{u}}$ on $\partial\Omega_u$ for the mechanical problem; prescribed fluid flux $\bar{\boldsymbol{q}}$ on $\partial\Omega_q$ and prescribed pressure $\bar{p}$ on $\partial\Omega_p$ for the fluid problem, with $\partial\Omega_t\cup\partial\Omega_u = \partial\Omega_q\cup\partial\Omega_p = \partial\Omega$.
        \textit{Middle}: RVE$_\mathrm{c}$ at the mesoscale, consisting of penny-shaped fluid-filled mesocrack inclusions with average pressure $p_\mathrm{c}$ embedded in a porous matrix with stiffness $\mathbb{C}_\mathrm{m}$.
        \textit{Right}: RVE$_\mathrm{p}$ at the microscale, consisting of spherical micropore inclusions with average pressure $p_\mathrm{p}$ distributed in a solid matrix with stiffness $\mathbb{C}_\mathrm{s}$.}
		\label{fig:scale-seperation}
	\end{figure}

We consider a solid body $\Omega \subset \mathbb{R}^{n_\mrm{dim}}$ $\left(n_{\mrm{dim}} = 2,3\right)$ embedded with a lower--dimensional macroscopic fracture $\Gamma \subset \mathbb{R}^{n_{\mrm{dim}-1}}$ which is regularized by the well-known phase-field method into $\Gamma_{\ell} \subset \mathbb{R}^{n_{\mrm{dim}}}$. 
As shown in Fig.~\ref{fig:scale-seperation} (left), $v$ is the phase-field variable, $v \in [0,1]$, with $v=0$ representing the fractured state and $v=1$ representing the intact material.
The body is subjected to a surface force $\mbfs{\bar{t}}$ on boundary $\partial \Omega_t$ and a flux $\mbfs{\bar{q}}$ on $\partial \Omega_q$.  In addition, $\bar{p}$ and $\bar{\mbfs{u}}$ are the prescribed pressure and displacement on boundary $\partial \Omega_p$ and $\partial \Omega_u$, respectively. 

Additionally, we consider the material at the mesoscale, i.e., RVE$_\mathrm{c}$ in Fig.~\ref{fig:scale-seperation} (middle), which is composed of porous matrix (index $\mathrm{m}$, volume fraction $\varphi_\mathrm{m} = V_\mathrm{m}/V$) and randomly distributed mesocracks (index $\mathrm{c}$, volume fraction $\varphi_\mathrm{c}=V_\mathrm{c}/V$), and $\varphi_\mathrm{m}+\varphi_\mathrm{c}=1$, where $V$ is the volume of RVE$_\mathrm{c}$, $V_\mathrm{m}$ is the volume of the porous matrix and $V_\mathrm{c}$ is the total volume of mesocracks. 
At the lower microscale, i.e., RVE$_\mathrm{p}$ in Fig.~\ref{fig:scale-seperation} (right), the porous material is made up of solid matrix (index $\mathrm{s}$, volume fraction $\varphi_\mathrm{s}=V_\mathrm{s}/V_\mathrm{m}$) and micropores (index $\mathrm{p}$, volume fraction $\varphi_\mathrm{p}=V_\mathrm{p}/V_\mathrm{m}$), and $\varphi_\mathrm{s}+\varphi_\mathrm{p}=1$, where $V_\mathrm{s}$ and $V_\mathrm{p}$ are the volume of solid matrix and the total volume of micropores. 
Note that the aforementioned definitions of $\varphi_\mathrm{m}$, $\varphi_\mathrm{c}$, $\varphi_\mathrm{s}$, and $\varphi_\mathrm{p}$ are Eulerian, which represent the volume fraction of the corresponding microstructures in the current configuration.
Using the reference volumes $V^0$ and $V_\mrm{m}^0$, the Lagrangian porosities can be defined as $\phi_\mathrm{m} = V_\mathrm{m}/V^0$, $\phi_\mathrm{c} = V_\mathrm{c}/V^0$, $\phi_\mathrm{p} = V_\mathrm{p}/V_\mathrm{m}^0$, and $\phi_\mathrm{s}=V_\mathrm{s}/V_\mathrm{m}^0$.

The fluids in the mesocracks and micropores are interconnected through a permeable matrix, allowing fluid mass exchange between the two pore systems.
However, the average fluid pressures in the mesocracks (denoted as $p_\mathrm{c}$) and micropores (denoted as $p_\mathrm{p}$) may not always be equilibrated i.e., $p_\mathrm{c} \neq p_\mathrm{p}$ if the matrix permeability $K_p$ is small, which is often the case. 
In the following, we will derive the macroscopic state equations of this two-scale double porosity medium. Unlike our previous work in \cite{you2023onporo}, the consideration of $p_\mathrm{c} \neq p_\mathrm{p}$ allows a more general analysis of fluid flow and solid deformation at the two different scales. For example, when viscous stress dominates the action of the fluid, there will be a pressure jump between the fracture and the porous matrix. This pressure jump will enhance mass transfer between the two pore regions, thereby affecting their mechanical behavior.
	
Recalling the microporomechanics model of \cite{pichler2010cracking} and considering the fluid saturated condition, the macroscopic state equations of this two-scale double-porosity media are defined as
\begin{equation}
  \boldsymbol{\sigma} =  \mathbb{C}_\mrm{eff}: \mbfs{\eps}(\mbfs{u}) -  \boldsymbol{\alpha}_\mathrm{c} p_\mathrm{c} - \boldsymbol{\alpha}_\mathrm{p} p_\mathrm{p}  
  ,
  \label{eq:total stress}
\end{equation}
where $\mathbb{C}_\mathrm{eff}$ denotes the effective elastic tensor, $\boldsymbol{\varepsilon}$ is the macroscopic strain defined as $\boldsymbol{\varepsilon} = \nabla^\mathrm{sym}\boldsymbol{u}$, $\boldsymbol{\alpha}_\mathrm{c}$ and $\boldsymbol{\alpha}_\mathrm{p}$ are Biot coefficient tensors at different scales respectively\footnote{At the pressure equilibrium between the two scales (i.e., $p_\mrm{c} = p_\mrm{p}=p$), Eq.~\eqref{eq:total stress} reduces to a familiar form:
\begin{equation}
  \boldsymbol{\sigma} =  \mathbb{C}_\mrm{eff}: \mbfs{\eps}(\mbfs{u}) -  \boldsymbol{\alpha} p
  ,
  \label{eq:total_stress_single}
\end{equation}
with $\boldsymbol{\alpha}=\boldsymbol{\alpha}_\mrm{c} + \boldsymbol{\alpha}_\mrm{p}$ being the Biot coefficient for the single continuum system.}.

The effective stiffness tensor in Eq.~\eqref{eq:total stress} accounts for the presence of mesocracks and micropores, which can be defined using volume averaging as
\begin{equation}
    \mathbb{C}_\mrm{eff} := (1-\varphi_\mathrm{c})\mathbb{C}_\mrm{m}:\mathbb{\bar{A}}_\mrm{m}
	,
    \label{eq:chom_general}
\end{equation}
where $\mathbb{\bar{A}}_\mathrm{m}$ is the average strain concentration tensor of the porous media in REV$_\mathrm{c}$. 
Similarly, with the volume fraction of micro-pore, $\varphi_\mathrm{p}$, and the average strain concentration tensor of the solid matrix in REV$_\mathrm{p}$, $\mathbb{\bar{A}}_\mathrm{s}$, we can define the homogenized stiffness tensor of  REV$_\mathrm{p}$, i.e., $\mathbb{C}_\mathrm{m}$ as
\begin{equation}
    \mathbb{C}_\mathrm{m} : =
    (1-\varphi_\mathrm{p})\mathbb{C}_\mathrm{s}:\mathbb{\bar{A}}_\mathrm{s}
    .
	\label{eq:chom_m}
\end{equation}
In this study, $\mathbb{C}_\mathrm{m}$ is considered to be an isotropic fourth-order tensor and thus can be written as
\begin{equation}
    \mathbb{C}_\mathrm{m}= 3k_\mrm{m} \mathbb{J}+2\mu_\mrm{m} \mathbb{K} 
	\label{eq:chom_m_iso}
\end{equation}
with $k_\mathrm{m}$ and $\mu_\mathrm{m}$ the bulk and shear moduli of the porous media.

Furthermore, the Biot coefficients $\boldsymbol{\alpha}_\mathrm{c}$ and $\boldsymbol{\alpha}_\mathrm{p}$ in Eq. \eqref{eq:total stress} are given as~\citep{dormieux2006microporomechanics}
	\begin{align}
		\label{eq:alpha_c}
		\mbfs{\alpha}_\mathrm{c} & = \mbfs{\delta}:\left( \mathbb{I} - (1-\varphi_\mathrm{c})\mathbb{\bar{A}}_\mrm{m}\right)
		,
		\\
		\label{eq:alpha_p}
		\mbfs{\alpha}_\mathrm{p} & = (1-\varphi_\mathrm{c}) \mbfs{\alpha}_\mrm{m}:\mathbb{\bar{A}}_\mrm{m}
		,
	\end{align}
	where $\mbfs{\alpha}_\mrm{m}$ is the isotropic Biot coefficient tensor for the porous matrix given by
	\begin{equation}
		\mbfs{\alpha}_\mathrm{m} = \mbfs{\delta}:\left( \mathbb{I} - (1-\varphi_\mathrm{p})\mathbb{\bar{A}}_\mathrm{s}\right) 
		\label{eq:alpha_m}
		.
	\end{equation}
 We note that $\mbfs{\alpha}_\mathrm{p} \neq \mbfs{\alpha}_\mrm{m}$ as $\mbfs{\alpha}_\mathrm{p}$ is defined throughout the domain $\Omega$ taking into account for the state of the meso-cracks, while $\mbfs{\alpha}_\mathrm{m}$ is defined in the domain excluding the mesocracks.

By denoting the compliance tensor of $\mathbb{C}_\mrm{s}$ as $\mathbb{S}_\mrm{s}$ (i.e., $\mathbb{S}_\mrm{s} = \left(\mathbb{C}_\mrm{s}\right)^{-1}$) and multiplying Eq.~\eqref{eq:chom_m} by $\mathbb{S}_\mathrm{s}$, we obtain 
 	\begin{equation}
(1-\varphi_\mathrm{p})\mathbb{\bar{A}}_\mathrm{s} = \mathbb{S}_\mathrm{s}:\mathbb{C}_\mathrm{m}
		.
        \label{eq:intermediate_alpha_m}
	\end{equation}
Substituting Eq.~\eqref{eq:intermediate_alpha_m} into Eq.~\eqref{eq:alpha_m} yields the Biot coefficient for porous media as
 \begin{equation}
    \mbfs{\alpha}_\mathrm{m} = \mbfs{\delta}:\left(\mathbb{I} - \mathbb{S}_\mathrm{s}:\mathbb{C}_\mathrm{m} \right) = \left( 1-\frac{k_\mathrm{m}}{k_\mathrm{s}}\right)\boldsymbol{\delta} = \alpha_\mathrm{m} \boldsymbol{\delta}
    ,
    \label{eq:alpha_m_final}
\end{equation}
where $k_\mathrm{s}$ denotes the bulk modulus of solid matrix and $\alpha_m$ is the Biot coefficient for the porous media.

Similarly, combining Eq. \eqref{eq:chom_general} with Eqs.~\eqref{eq:alpha_c} and \eqref{eq:alpha_p}, we have
 \begin{align}
     \label{eq:alpha_c_final}
		\mbfs{\alpha}_\mathrm{c} & = \mbfs{\delta}:\left(\mathbb{I} - \mathbb{S}_\mathrm{m}:\mathbb{C}_\mathrm{eff} \right)= \boldsymbol{\delta} - \frac{\boldsymbol{\delta}:\mathbb{C}_\mathrm{eff}}{3k_\mathrm{m}}
		\\
		\label{eq:alpha_p_final}
		\mbfs{\alpha}_\mathrm{p} & = \mbfs{\alpha}_\mrm{m}:\mathbb{S}_\mathrm{m}:\mathbb{C}_\mathrm{eff} = \alpha_\mathrm{m} \frac{\boldsymbol{\delta}:\mathbb{C}_\mathrm{eff}}{3k_\mathrm{m}}
        ,
 \end{align}
 where $\mathbb{S}_\mrm{m}$ is the compliance tensor of $\mathbb{C}_\mrm{m}$ defined as $\mathbb{S}_\mrm{m} = \left(\mathbb{C}_\mrm{m}\right)^{-1}$. 
 Note that the effective stiffness $\mathbb{C}_\mathrm{eff}$ may or may not be 
 isotropic depending on the energy decomposition scheme used in the phase-field model. 
 \cite{you2023onporo} applied the volumetric--deviatoric energy decomposition~\citep{Amor2009}, which results in an isotropic $\mathbb{C}_\mathrm{eff}$ and thus an isotropic Biot coefficient.
 In this study, however, we consider a more general case where $\mathbb{C}_\mathrm{eff}$ may be anisotropic, and the Biot coefficients $\boldsymbol{\alpha}_\mathrm{c}$ and $\boldsymbol{\alpha}_\mathrm{p}$ may consequently be anisotropic tensors. 

The fluid content changes in mesoscale (RVE$_\mathrm{c}$) and microscale (RVE$_\mathrm{p}$) can be defined as
 \begin{align}
    \zeta_\mathrm{c} := \phi_\mathrm{c}-\phi^0_\mathrm{c} \label{eq:porosity variation M}\\
    \zeta_\mathrm{p}:= \phi_\mathrm{p}-\phi^0_\mathrm{p} \label{eq:porosity variation m}
    .
 \end{align} 

\begin{remark}
$\zeta_c$ and $\zeta_p$ in Eqs.~\eqref{eq:porosity variation M} and \eqref{eq:porosity variation m} are defined using the Lagrangian description of the porosities of the mesocracks ($\phi_\mathrm{c}$) and micropores ($\phi_\mathrm{p}$), which describe the volume changes relative to the reference volumes. Under small deformations, the Lagrangian and Eulerian descriptions ($\varphi_\mathrm{c}$ and $\varphi_\mathrm{p}$) can be assumed to be approximately equal, but their variations are not identical. For example, the rate of $\phi_\mathrm{c}$ is defined as $\dot{\phi}_\mathrm{c} = \varphi_\mathrm{c} \dot{\boldsymbol{\varepsilon}}_\mathrm{c}$ with $\dot{\boldsymbol{\varepsilon}}_\mathrm{c}$ being the crack strain rate~\citep{pichler2010cracking}.
\end{remark}

Additionally, the state equations for the mesocracks and micropores under a fully saturated condition are given by~\citep{dormieux2006microporomechanics}
 \begin{align}
    \zeta_\mathrm{c} = \boldsymbol{\alpha}_\mathrm{c}:\mbfs{\eps}(\mbfs{u})+ \frac{p_\mathrm{c}}{N_\mathrm{cc}} + \frac{p_\mathrm{p}}{N_\mathrm{cp}} \label{eq:zeta_c}\\
    \zeta_\mathrm{p} = \boldsymbol{\alpha}_\mathrm{p}:\mbfs{\eps}(\mbfs{u})+ \frac{p_\mathrm{c}}{N_\mathrm{pc}} + \frac{p_\mathrm{p}}{N_\mathrm{pp}} \label{eq:zeta_p}
 \end{align}
 with 
 \begin{align}
		\begin{cases}
  \dfrac{1}{N_\mathrm{cc}} = \mbfs{\delta}:\mathbb{S}_\mrm{m}:\left(\mbfs{\alpha}_\mathrm{c} - \varphi_\mathrm{c} \mbfs{\delta}\right)  \\
			\dfrac{1}{N_\mathrm{cp}} = \mbfs{\delta}:\mathbb{S}_\mrm{m}:\left[\mbfs{\alpha}_\mathrm{p} - (1-\varphi_\mathrm{c}) \mbfs{\alpha}_\mrm{m}\right] \\
			\dfrac{1}{N_\mathrm{pc}} =  \mbfs{\alpha}_\mrm{m}:\mathbb{S}_\mrm{m}:\left(\varphi_\mathrm{c}\mbfs{\delta} - \mbfs{\alpha}_\mathrm{c}\right)  \\
			\dfrac{1}{N_\mathrm{pp}} = \mbfs{\alpha}_\mrm{m}:\mathbb{S}_\mrm{m}:\left[(1-\varphi_\mathrm{c})\mbfs{\alpha}_\mrm{m} - \mbfs{\alpha}_\mathrm{p}\right] + \dfrac{1 - \varphi_\mathrm{c}}{N_\mrm{m}}
		\end{cases}
		\label{eq:M_matrix}
	\end{align}
	where $N_\mrm{m}$ is the Biot modulus of the porous media defined as
 \begin{equation}
     \frac{1}{N_\mathrm{m}} = \left(\boldsymbol{\alpha}_m - \varphi_\mathrm{p}\boldsymbol{\delta}\right):\mathbb{S}_\mathrm{s}:\boldsymbol{\delta} = \frac{\alpha_\mathrm{m}-\varphi_\mathrm{p}}{k_\mathrm{s}}
     .
     \label{eq:N_m}
 \end{equation}

Combining Eqs. \eqref{eq:alpha_c_final}, \eqref{eq:alpha_p_final}, \eqref{eq:alpha_m_final} and \eqref{eq:N_m}, the Biot moduli \eqref{eq:M_matrix} become 
\begin{align}
		\begin{cases}
  \dfrac{1}{N_\mathrm{cc}} = \dfrac{1- \frac{k_\mathrm{eff}}{k_\mathrm{m}} - \varphi_\mathrm{c}}{k_\mrm{m}} \\
			\dfrac{1}{N_\mathrm{cp}} = \dfrac{1}{N_\mathrm{pc}} = \dfrac{\alpha_\mrm{m}\left[\varphi_\mathrm{c}- \left(1-\frac{k_\mathrm{eff}}{k_\mathrm{m}}\right)\right]}{k_\mrm{m}} \\[6pt]
			\dfrac{1}{N_\mathrm{pp}} = \dfrac{\alpha_\mrm{m}^2(1 - \frac{k_\mathrm{eff}}{k_\mathrm{m}} - \varphi_\mathrm{c})}{k_\mrm{m}} + \dfrac{(1-\varphi_\mathrm{c})(\alpha_\mrm{m}-\varphi_\mrm{p})}{k_\mrm{s}}
		\end{cases}
		\label{eq:M_matrix——2}
\end{align}
with $k_\mathrm{eff}$ defined as
\begin{equation}
    k_\mathrm{eff} = \frac{1}{9}\boldsymbol{\delta}:\mathbb{C}_\mathrm{eff}:\boldsymbol{\delta}
\label{eq:bulk modulus}    
\end{equation}

Eq.~\eqref{eq:M_matrix——2} shows that the Biot moduli are symmetric, i.e., $\frac{1}{N_\mathrm{cp}} = \frac{1}{N_\mathrm{pc}}$. 
Since the pressure within a given pore network must promote, not impede, its own volumetric expansion, the diagonal Biot moduli must be non-negative, i.e., $\frac{1}{N_\mathrm{cc}}\geq 0$ and $\frac{1}{N_\mathrm{pp}} \geq 0$. 
These constraints yield $\frac{1}{N_\mathrm{pc}}=\frac{1}{N_\mathrm{cp}}\leq 0$, signifying a negative cross-coupling between the two pore pressures, i.e., an increase in $p_\mathrm{c}$ diminishes $\phi_\mathrm{p}$, and vice versa. This behavior admits a clear physical interpretation. At a constant total porosity (i.e., in the absence of microstructural growth), an elevated $p_\mathrm{c}$ drives the dilation of mesocracks, which must be accommodated by a commensurate compression of the micropore space.





\subsection{A phase-field fracture model for dual-continuum porous media}
Following the variational approach to fracture proposed by~\cite{Francfort1998}, the total energy of the system is defined as a sum of the Helmholtz free energy, $\psi$, and the surface energy:

\begin{equation}
\mathcal{E} (\boldsymbol{\varepsilon},\Gamma) : = \int_{\Omega\setminus\Gamma} \psi (\boldsymbol{\varepsilon})  \, \mathd V      
+ \int_\Gamma \Gc \, \mathd S
,
\label{eq:FM_total_energy}
\end{equation}
where $G_c$ is the critical energy release rate.

The Helmholtz free energy, $\psi$, considering the fluid-filled dual continuum, can be derived as
(see Appendix \ref{ap:free energy} for the details)
\begin{equation}
    \psi(\boldsymbol{\varepsilon}, \zeta_\mathrm{c}, \zeta_\mathrm{p}) = \underbrace{\frac{1}{2} \boldsymbol{\varepsilon}(\boldsymbol{u}):\mathbb{C}_\mathrm{eff}:\boldsymbol{\varepsilon}(\boldsymbol{u})}_{\psi_e(\boldsymbol{\varepsilon})} + \underbrace{\frac{p^2_\mathrm{c}}{2N_\mathrm{cc}} + \frac{p^2_\mathrm{p}}{2N_\mathrm{pp}} + \frac{p_\mathrm{c} p_\mathrm{p}}{N_\mathrm{pc}}}_{\psi_f(\boldsymbol{\varepsilon}, \zeta_c, \zeta_p)}
    ,
    \label{eq:psi}
\end{equation}
where $\psi_e(\boldsymbol{\varepsilon})$ denotes the strain energy density and $\psi_f(\boldsymbol{\varepsilon}, \zeta_c, \zeta_p)$ accounts for the interaction between the mesocrack pressure and micropore pressure. 
The control variables $\zeta_\mathrm{c}$ and $\zeta_\mathrm{p}$ are implicit in the quantities $p_\mathrm{c}$ and $p_\mathrm{p}$, according to Eqs.~\eqref{eq:porosity variation M} and~\eqref{eq:porosity variation m}. Therefore, we can obtain the following constitutive relations as
\begin{equation}
    \boldsymbol{\sigma} = \frac{\partial \psi}{\partial \boldsymbol{\varepsilon}},
    \,
    p_\mathrm{c} = \frac{\partial \psi}{\partial \zeta_\mathrm{c}},
    \,
    p_\mathrm{p} = \frac{\partial \psi}{\partial \zeta_\mathrm{p}}.
\end{equation}

\begin{remark}
    Note that the derivation of Eq. \eqref{eq:psi} is based on the grand-canonical potential \citep{anand20152014}, following from the Legendre transform 
\begin{equation*}
   \psi^*(\boldsymbol{\varepsilon},p_\mathrm{c},p_\mathrm{p}) = \psi - p_\mathrm{c}\zeta_\mathrm{c} - p_\mathrm{p}\zeta_\mathrm{p}.
   \label{eq:psi_*}
\end{equation*}
With this, by substituting Eqs.~\eqref{eq:porosity variation M} and~\eqref{eq:porosity variation m}, we have 
\begin{equation}
    \boldsymbol{\sigma} = \frac{\partial \psi^*}{\partial \boldsymbol{\varepsilon}}, \,
    \zeta_\mathrm{c} = - \frac{\partial \psi^*}{\partial p_\mathrm{c}}
    ,\,
    \zeta_\mathrm{p} = - \frac{\partial \psi^*}{\partial p_\mathrm{p}}
    .
    \label{eq:constitutive relations}
\end{equation}
This procedure was followed by several existing works \citep{dormieux2006micromechanical,pichler2010cracking,xie2012micromechanical} on microporoelasticity models. \cite{zhu2023microporomechanics} used a slightly different Legendre transform process to obtain the Gibbs free energy, which however leads to the same relations as Eq.~\eqref{eq:constitutive relations}.
\end{remark}

Substituting Eq.~\eqref{eq:psi} into Eq.~\eqref{eq:FM_total_energy} yields

\begin{equation}
\mathcal{E} (\boldsymbol{\varepsilon},\zeta_c,\zeta_p,\Gamma)  = \int_{\Omega\setminus\Gamma} \psi_e (\boldsymbol{\varepsilon})  \, \mathd V    
+\int_{\Omega\setminus\Gamma} \psi_f (\boldsymbol{\varepsilon},\zeta_c,\zeta_p)  \, \mathd V    
+ \int_\Gamma \Gc \, \mathd S
.
\label{eq:FM_total_energy_DP}
\end{equation}
Following the now-standard regularization proposed by~\cite{Bourdin2000}, we define the regularized total energy of Eq.~\eqref{eq:FM_total_energy_DP} as
\begin{equation}
\mathcal{E}_\ell (\boldsymbol{\varepsilon},,\zeta_c,\zeta_p,v) : = \int_{\Omega} \psi_e (\boldsymbol{\varepsilon},v)  \, \mathd V    
+\int_{\Omega} \psi_f (\boldsymbol{\varepsilon},\zeta_c,\zeta_p,v)  \, \mathd V    
+ \int_\Omega \frac{\Gc}{4c_n} \left[\frac{(1-v)^n}{\ell}+ \ell \nabla v\cdot \nabla v \right] \, \mathd V
.
\label{eq:FM_total_energy_DP}
\end{equation}
where $\ell$ is the length scale parameter and $c_n$ is a normalization constant defined as $c_n = \int_0^1 (1-s)^n \mathd s$.

With the energy decomposition~\citep{Amor2009, Freddi2010}, the regularized strain energy is written as
\begin{equation}
    \psi_e(\boldsymbol{\varepsilon},v) = g(v) \psi^+_e(\boldsymbol{\varepsilon}) + \psi^-_e(\boldsymbol{\varepsilon}),
\end{equation} 
where $\psi_e^+$ and $\psi_e^-$ are the positive and negative strain energies, and $g(v)$ is the degradation function, defined as
\begin{equation}
    g(v) = (1-\kappa)v^2 + \kappa
    ,
\end{equation}
with $\kappa = 10^{-8}$ throughout this study.

Thus far, we have not specified the expression of $\mathbb{C}_\mathrm{eff}$, which accounts for the presence of mesocracks and micropores. 
Following our previous work \citep{you2023onporo}, we employ the phase-field model to capture fractures and the expression of $\mathbb{C}_\mathrm{eff}$ can be specified through a phase-field degraded strain energy density:
\begin{equation}
   \mathbb{C}_\mathrm{eff} (v) = \frac{\partial^2 \psi_e}{\partial \boldsymbol{\varepsilon}^2} = g(v)\frac{\partial^2 \psi_e^+}{\partial \boldsymbol{\varepsilon}^2} + \frac{\partial^2 \psi_e^-}{\partial \boldsymbol{\varepsilon}^2}
   .
   \label{eq:Ceff_strain_energy}
\end{equation}
The phase-field variable ($v$) dependent $\mathbb{C}_\mathrm{eff}$ captures the dependency of the Biot moduli on $v$ through Eqs.~\eqref{eq:M_matrix——2}, and thus we simply have
\begin{equation}
\psi_f (\boldsymbol{\varepsilon},\zeta_c,\zeta_p,v)
= \frac{p^2_\mathrm{c}}{2N_\mathrm{cc}(v)} + \frac{p^2_\mathrm{p}}{2N_\mathrm{pp}(v)} + \frac{p_\mathrm{c} p_\mathrm{p}}{N_\mathrm{pc}(v)} 
\end{equation}


The momentum balance and the phase-field evolution equations can be obtained by taking the first variation of the functional $\mathcal{E}_\ell (\mbfs{\varepsilon},v)$ with respect to $\bm{\varepsilon}$ and $v$:
\begin{align}
    \label{eq: displacement}
    &\nabla \cdot \left[ \mathbb{C}_\mrm{eff} : \mbfs{\varepsilon}(\mbfs{u}) -  {\boldsymbol{\alpha}_\mathrm{c}} p_\mathrm{c} -  {\boldsymbol{\alpha}_\mathrm{p}} p_\mathrm{p}  \right]  + \mbfs{b} = \mathbf{0} \\
    \label{eq: phase-field}
    &2(1-\kappa)v\mathcal{D}  + \frac{G_c}{4c_n} \left( -\frac{n (1-v)^{n-1}}{\ell} + 2\ell \Delta v\right) = 0
\end{align}
with boundary conditions $\mbfs{\sigma}\cdot n=\bar{\mbfs{t}}$ on $\partial \Omega_t$ and $\nabla v \cdot \mbfs{n} =0 $ on $\partial \Omega$. 

The term $\mathcal{D}$ in Eq. \eqref{eq: phase-field} is the phase-field driving force, which is defined as
\begin{equation}
    \mathcal{D} = \psi_e^+(\boldsymbol{u}) + \frac{p^2_\mathrm{c}}{2}\frac{\partial (1/N_\mathrm{cc})}{\partial v} + \frac{p^2_\mathrm{p}}{2}\frac{\partial (1/N_\mathrm{pp})}{\partial v} + p_\mathrm{c}p_\mathrm{p}\frac{\partial (1/N_\mathrm{pc})}{\partial v}
    \label{eq:D}
\end{equation}
where the phase-field derivatives of the Biot moduli, obtained by differentiating Eq. \eqref{eq:M_matrix——2} with respect to $v$, are
\begin{align}
    \frac{\partial (1/N_\mathrm{cc})}{\partial v} &= -\frac{1}{k_\mrm{m}^2}\frac{\partial k_\mathrm{eff}}{\partial v} - \cancel{\frac{1}{k_\mrm{m}}\frac{\partial \varphi_\mathrm{c}}{\partial v}}, \label{eq:dNcc} \\
    \frac{\partial (1/N_\mathrm{pc})}{\partial v} &= -\alpha_\mathrm{m} \frac{\partial (1/N_\mathrm{cc})}{\partial v}, \label{eq:dNpc} \\
    \frac{\partial (1/N_\mathrm{pp})}{\partial v} &= \alpha_\mathrm{m}^2 \frac{\partial (1/N_\mathrm{cc})}{\partial v} - \cancel{\frac{\alpha_\mathrm{m} - \varphi_\mathrm{p}}{k_\mrm{s}}\frac{\partial \varphi_\mathrm{c}}{\partial v}}. \label{eq:dNpp}
\end{align}
Eqs. \eqref{eq:dNcc}--\eqref{eq:dNpp} imply that $\varphi_\mathrm{p}$ is invariant with the damage ($v$), which means the micropores do not evolve with damage, and the volume fraction of mesocracks $\varphi_\mathrm{c}$ is assumed to be negligible as we consider the penny--shaped cracks with zero volume, unlike $v$-dependent porosity models~\citep{yi2020consistent,you2023onporo}. 
This is consistent with the physical picture that the storage of the material is mainly due to micropores and will not be largely altered by the mesocracks.

For the energy decomposition in this study, we apply the no-tension model~\citep{Freddi2010}, where the positive strain ($\bm{\eps}^+$) is a positive definite tensor and is coaxial with $\bm{\eps}$.
Denoting the principal strains with $\varepsilon_i$ with $i \in\{1,2,3\}$, we have
\begin{equation}\label{Eq:mas-eps}
    \begin{aligned}
        \bm{\varepsilon} = \overset{3}{\underset{i=1}{\sum}}\varepsilon_i \bm{n}_{(i)} \otimes \bm{n}_{(i)} := \varepsilon_i \bm{M}_i,
    \end{aligned}
\end{equation}
where $\bm{n}_{(i)}$ is the eigenvector and parentheses around an index indicate that no sum is taken, i.e., the usual summation convention is suspended.
Then the positive and the negative strains are given by
\begin{equation}\label{Eq:mas-epspn}
    \begin{aligned}
        \bm{\varepsilon}_+ =  a_i \bm{M}_i, \quad \bm{\varepsilon}_- = \bm{\varepsilon} - \bm{\varepsilon_+} .
    \end{aligned}
\end{equation}
The positive definiteness of $\bm{\varepsilon}_+$ depends not only on the sign of $\varepsilon_i$ but also on the material's Poisson's ratio.
For the detailed construction of $\bm{\varepsilon}_+$, we refer to \cite{Freddi2010} and the references therein.
Using the decomposed strains, the positive and negative strain energies are given as
\begin{equation}
    \psi^+_e(\boldsymbol{\varepsilon})  = \frac{1}{2} \left[\mathrm{Tr}(\boldsymbol{\varepsilon}_+)\right]^2 + \mu\boldsymbol{\varepsilon}_+ : \boldsymbol{\varepsilon}_+ , \,
    \psi^-_e(\boldsymbol{\varepsilon})  = \frac{1}{2} \left[\mathrm{Tr}(\boldsymbol{\varepsilon}_-)\right]^2 + \mu\boldsymbol{\varepsilon}_- : \boldsymbol{\varepsilon}_- .
\end{equation}
Note that with the no-tension strain energy decomposition, $\mathbb{C}_\mathrm{eff}$ is anisotropic.  

\subsection{Mass balance equations}
 The fluid mass balances in mesocracks and micropores can be written as\footnote{The fluid balance equation can also be derived from the mixed variational approach as presented in \cite{ulloa2022variational}, where a mixed fluid dissipation functional was introduced. Although only the single-porosity fluid equation was derived in \cite{ulloa2022variational}, an extension to double-porosity media will be a promising work in future.}
\begin{align}
	\label{eq:mass_balance}
		\dfrac{\partial (\rho_f \phi_\mathrm{c})}{\partial t} + \rho_f \nabla \cdot \mbf{v}_\mathrm{c} = Q_\mathrm{c} + r_\mathrm{c} & \quad \mrm{in}\quad \Omega 
		\\
  		\dfrac{\partial (\rho_f \phi_\mathrm{p})}{\partial t} + \rho_f \nabla \cdot \mbf{v}_\mathrm{p} = Q_\mathrm{p} +  r_\mathrm{p} & \quad \mrm{in}\quad \Omega_\Gamma 
\end{align}
where $\rho_f$ is the fluid density. $Q_\mathrm{c}$ and $Q_\mathrm{p}$ are the source or sink terms, $\mbf{v}_\mathrm{c}$ and $\mbf{v}_\mathrm{p}$ are the fluid velocities in the mesocracks and micropores, respectively. We consider Darcy's flow at both length scales that are given by 
\begin{align}
    \mbf{v}_\mathrm{c}  = - \frac{K_\mathrm{c}}{\mu_f}(\nabla p_\mathrm{c}-\rho_f \boldsymbol{g}) \label{eq:darcy M},\\
    \mbf{v}_\mathrm{p}  = - \frac{K_\mathrm{p}}{\mu_f}(\nabla p_\mathrm{p}-\rho_f \boldsymbol{g}),
    \label{eq:darcy m}
\end{align}
with $K_\mathrm{c}$ the permeability for the fluid flow at the mesoscale, $K_\mathrm{p}$ the permeability for the fluid flow at the microscale and $\boldsymbol{g}$ the gravity. The permeability $K_\mrm{c}$ will be enhanced when the macroscopic fracture appears. We employ a volumetric flux equivalence, in which the permeability $K_\mathrm{c}$ is defined as
\begin{equation}
    K_\mrm{c} =   \frac{w^3}{12h}\left(\boldsymbol{\delta} - \mathbf{n}_\Gamma \otimes \mathbf{n}_\Gamma\right),
    \label{eq: Kc}
\end{equation}
with $h$ being the characteristic element size, $w$ the fracture aperture, and $\mathbf{n}_\Gamma$ the normal direction of the fracture. The normal direction is estimated by the eigen vector of the maximum principal strain, and the fracture aperture $w$ is related to the maximum principal strain $\varepsilon_1$ and the characteristic element size by $h$ as~\citep{you2023onporo}
\begin{equation}
    w = \varepsilon_1 h.
    \label{eq:aperture}
\end{equation}

Substituting Eqs. \eqref{eq:porosity variation M}, \eqref{eq:porosity variation m}, \eqref{eq:darcy M} and \eqref{eq:darcy m} into Eqs. \eqref{eq:mass_balance}, we have
\begin{align}
    \label{eq:mass_balance_c}
    &\frac{\partial}{\partial t} \left[ \rho_f \left( \boldsymbol{\alpha}_\mathrm{c}: \boldsymbol{\varepsilon} + \frac{p_\mathrm{c}}{N_\mathrm{cc}} + \frac{p_\mathrm{p}}{N_\mathrm{cp}} \right) \right] - \nabla \cdot \left( \frac{\rho_f K_\mathrm{c}}{\mu}(\nabla p_\mathrm{c}-\rho_f \boldsymbol{g})\right) = Q_\mathrm{c} + r_\mathrm{c}  \\
    \label{eq:mass_balance_p}
    &\frac{\partial}{\partial t} \left[ \rho_f \left( \boldsymbol{\alpha}_\mathrm{p}: \boldsymbol{\varepsilon} + \frac{p_\mathrm{c}}{N_\mathrm{pc}} + \frac{p_\mathrm{p}}{N_\mathrm{pp}} \right) \right] - \nabla \cdot \left( \frac{\rho_f K_\mathrm{p}}{\mu}(\nabla p_\mathrm{p}-\rho_f \boldsymbol{g})\right) = Q_\mathrm{p} + r_\mathrm{p} 
\end{align}     

Furthermore, $r_\mathrm{c}$ and $r_\mathrm{p}$ are the terms of mass interaction between the two porous scales and satisfy the closure condition $r_\mathrm{c}+r_\mathrm{p}=0$. We can approximate the mass transfer as~\citep{kazemi1976numerical} 
\begin{equation}
     r_\mathrm{c} = -r_\mathrm{p}=\dfrac{\rho_f K_\mathrm{m}}{\mu} \mathcal{S} \dfrac{V}{h} \dfrac{2(p_\mathrm{p} - p_\mathrm{c})}{h/2}
\end{equation}
which automatically ensures the closure condition. Here, $\mathcal{S}$ is a shape factor which depends on the geometry of the mesocracks and micropores.

\section{Solution schemes} \label{sec:solution}
This section is devoted to numerical solution of the proposed coupling model, which involves four independent fields, i.e., the displacement $\boldsymbol{u}$, the fluid pressure in mesocracks $p_\mathrm{c}$, the fluid pressure in micropores $p_\mathrm{p}$ and the phase-field $v$. We adopt the staggered scheme to solve the phase-field problem, which is widely used in phase-field fracture modeling~\citep{Bourdin2000,miehe2010phase,ambati2015review}. In each time step, we first solve the coupled $p_\mathrm{c}-p_\mathrm{p}-\boldsymbol{u}$ problem with a fixed phase-field variable $v$, and then update the phase-field variable by solving the phase-field equation with the updated displacement and pore pressures. The four governing equations are solved using the finite element method, and the time integration is performed using the backward Euler scheme.

\subsection{Fixed stress split for the double-porosity problem}
We adopt the fixed stress split scheme \citep{kim2011stability,you2023onporo} to decouple the mechanics from the flow problems, and solve them in a staggered manner. 
The key idea of the fixed stress split scheme is to fix the volumetric stress during the iteration in each time step, which has been shown to be unconditionally stable for single-porosity poromechanics problems \cite{kim2011stability} with lower-order elements. 
Here, this scheme needs to be extended to the double-porosity hydromechanical coupling problem.

At the $p_\mathrm{c}-p_\mathrm{p}-\boldsymbol{u}$ iteration, $v$ is frozen with the staggered scheme. 
According to Eqs.~\eqref{eq:porosity variation M} and \eqref{eq:porosity variation m}, we have
\begin{align}
\label{eq:variation_phic}
    \frac{\partial \phi_\mathrm{c}}{\partial t}  = \boldsymbol{\alpha}_\mathrm{c}:\frac{\partial \boldsymbol{\varepsilon}}{\partial t} + \frac{1}{N_\mathrm{cc}}\frac{\partial p_\mathrm{c}}{\partial t} + \frac{1}{N_\mathrm{cp}}\frac{\partial p_\mathrm{p}}{\partial t} \\
    \label{eq:variation_phip}
    \frac{\partial \phi_\mathrm{p}}{\partial t}  = \boldsymbol{\alpha}_\mathrm{p}:\frac{\partial \boldsymbol{\varepsilon}}{\partial t} + \frac{1}{N_{pc}}\frac{\partial p_\mathrm{c}}{\partial t} + \frac{1}{N_{pp}}\frac{\partial p_\mathrm{p}}{\partial t}
\end{align}
Applying the backward Euler scheme, Eqs.~\eqref{eq:variation_phic} and \eqref{eq:variation_phip} at time step $i$ and coupling iteration $j$ yield
\begin{align}
    \frac{\phi^{i,j}_\mathrm{c} - \phi^{i-1}_\mathrm{c}}{\Delta t}  
    = \frac{ \boldsymbol{\alpha}_\mathrm{c}:(\boldsymbol{\varepsilon}^{i,j} - \boldsymbol{\varepsilon}^{i-1} ) }{\Delta t} + \frac{1}{N_\mathrm{cc}}\frac{p^{i,j}_\mathrm{c} - p^{i-1}_\mathrm{c}}{\Delta t} + \frac{1}{N_\mathrm{cp}}\frac{p^{i,j}_\mathrm{p} - p^{i-1}_\mathrm{p}}{\Delta t} \label{eq:BE_c}\\
    \frac{\phi^{i,j}_\mathrm{p} - \phi^{i-1}_\mathrm{p}}{\Delta t}  
    = \frac{\boldsymbol{\alpha}_\mathrm{p}:( \boldsymbol{\varepsilon}^{i,j} - \boldsymbol{\varepsilon}^{i-1})}{\Delta t} + \frac{1}{N_{pc}}\frac{p^{i,j}_\mathrm{c} - p^{i-1}_\mathrm{c}}{\Delta t} + \frac{1}{N_{pp}}\frac{p^{i,j}_\mathrm{p} - p^{i-1}_\mathrm{p}}{\Delta t} \label{eq:BE_p}
\end{align}

In the fixed-stress scheme, the term, $\boldsymbol{\alpha}:\boldsymbol{\varepsilon}^{i,j}$, is expressed with the strain at the previous iteration, $\boldsymbol{\varepsilon}^{i,j-1}$ by freezing the compliance weighted volumetric total stress during the coupling iteration\footnote{For an isotropic $\mathbb{C}_\mrm{eff}$, this reduces to $\boldsymbol{\delta}:\boldsymbol{\sigma}^{i,j} = \boldsymbol{\delta}:\boldsymbol{\sigma}^{i,j-1}$, which is the identical form proposed in~\citep{kim2011stability}.}: 
$$ \mathbb{S}_\mathrm{eff}: \boldsymbol{\alpha} :\boldsymbol{\sigma}^{i,j} = \mathbb{S}_\mathrm{eff}: \boldsymbol{\alpha} :\boldsymbol{\sigma}^{i,j-1},$$ 
where $\mathbb{S}_\mathrm{eff}$ is the effective compliance tensor defined as the inverse of $\mathbb{C}_\mathrm{eff}$.
In our proposed double porosity system, we impose the following stress equalities for each scale individually:
\begin{align}
  \mathbb{S}_\mathrm{eff}: \boldsymbol{\alpha}_\mathrm{c} :\boldsymbol{\sigma}^{i,j} &
  = \mathbb{S}_\mathrm{eff}: \boldsymbol{\alpha}_\mathrm{c} :\boldsymbol{\sigma}^{i,j-1}, \label{eq:fixed stress split 1}\\
  \mathbb{S}_\mathrm{eff}: \boldsymbol{\alpha}_\mathrm{p}:\boldsymbol{\sigma}^{i,j} &= \mathbb{S}_\mathrm{eff}: \boldsymbol{\alpha}_\mathrm{p}:\boldsymbol{\sigma}^{i,j-1}, \label{eq:fixed stress split 2}
\end{align}
which can be simplified with Eqs.~\eqref{eq:alpha_c_final} and \eqref{eq:alpha_p_final} as
\begin{align}
  \left( \boldsymbol{\delta}:\mathbb{S}_\mathrm{eff} - \frac{\boldsymbol{\delta}}{3k_\mathrm{m}}\right):\boldsymbol{\sigma}^{i,j} &= \left( \boldsymbol{\delta}:\mathbb{S}_\mathrm{eff} - \frac{\boldsymbol{\delta}}{3k_\mathrm{m}}\right):\boldsymbol{\sigma}^{i,j-1}, \label{eq:fixed stress split 1-2}
  \\ \boldsymbol{\delta}:\boldsymbol{\sigma}^{i,j} &= \boldsymbol{\delta}:\boldsymbol{\sigma}^{i,j-1}. \label{eq:fixed stress split 2-2}
\end{align}
Eq.~\eqref{eq:fixed stress split 1-2} fixes a linear combination of stress components associated with the mesocrack Biot tensor, while Eq.~\eqref{eq:fixed stress split 2-2} independently fixes the mean stress $\boldsymbol{\delta}:\boldsymbol{\sigma}$. 
Note that for an isotropic $\mathbb{C}_\mathrm{eff}$, Eq.~\eqref{eq:fixed stress split 2-2} is a consequence of Eq.~\eqref{eq:fixed stress split 1-2}; for an anisotropic $\mathbb{C}_\mathrm{eff}$, they are independent.

Substituting the constitutive relation $\boldsymbol{\sigma} = \mathbb{C}_\mathrm{eff}:\boldsymbol{\varepsilon} - \boldsymbol{\alpha}_\mathrm{c} p_\mathrm{c}  - \boldsymbol{\alpha}_\mathrm{p} p_\mathrm{p}$ into Eq.~\eqref{eq:fixed stress split 1} and Eq.~\eqref{eq:fixed stress split 2} respectively, we have
\begin{multline}
\mathbb{S}_\mathrm{eff}: \boldsymbol{\alpha}_\mathrm{c} :\mathbb{C}_\mathrm{eff}:\boldsymbol{\varepsilon}^{i,j} = \mathbb{S}_\mathrm{eff}: \boldsymbol{\alpha}_\mathrm{c} :\mathbb{C}_\mathrm{eff}:\boldsymbol{\varepsilon}^{i,j-1} + \mathbb{S}_\mathrm{eff}: \boldsymbol{\alpha}_\mathrm{c} :\boldsymbol{\alpha}_\mathrm{c} \left(p^{i,j}_\mathrm{c} - p^{i,j-1}_\mathrm{c}\right) \\+ \mathbb{S}_\mathrm{eff}: \boldsymbol{\alpha}_\mathrm{c} :\boldsymbol{\alpha}_\mathrm{p} \left(p^{i,j}_\mathrm{p} - p^{i,j-1}_\mathrm{p}\right) ,
    \label{eq:fixed stress 2}
\end{multline}
\begin{multline}
    \mathbb{S}_\mathrm{eff}: \boldsymbol{\alpha}_\mathrm{p} :\mathbb{C}_\mathrm{eff}:\boldsymbol{\varepsilon}^{i,j} = \mathbb{S}_\mathrm{eff}: \boldsymbol{\alpha}_\mathrm{p} :\mathbb{C}_\mathrm{eff}:\boldsymbol{\varepsilon}^{i,j-1} + \mathbb{S}_\mathrm{eff}: \boldsymbol{\alpha}_\mathrm{p} :\boldsymbol{\alpha}_\mathrm{c} \left(p^{i,j}_\mathrm{c} - p^{i,j-1}_\mathrm{c}\right) \\+ \mathbb{S}_\mathrm{eff}: \boldsymbol{\alpha}_\mathrm{p} :\boldsymbol{\alpha}_\mathrm{p} \left(p^{i,j}_\mathrm{p} - p^{i,j-1}_\mathrm{p}\right) ,
    \label{eq:fixed stress p}
\end{multline}
which can be further simplified using Eq. \eqref{eq:alpha_c_final} and Eq. \eqref{eq:alpha_p_final} as (see Appendix~\ref{ap:fixed stress} for details)
\begin{align}
    \boldsymbol{\alpha}_\mathrm{c}:\boldsymbol{\varepsilon}^{i,j} &= \boldsymbol{\alpha}_\mathrm{c}:\boldsymbol{\varepsilon}^{i,j-1}
    + \boldsymbol{\alpha}_\mathrm{c}:\mathbb{S}_\mathrm{eff}:\boldsymbol{\alpha}_\mathrm{c}\left(p_\mathrm{c}^{i,j} - p_\mathrm{c}^{i,j-1}\right)
    + \boldsymbol{\alpha}_\mathrm{c}:\mathbb{S}_\mathrm{eff}:\boldsymbol{\alpha}_\mathrm{p}\left(p_\mathrm{p}^{i,j} - p_\mathrm{p}^{i,j-1}\right),
    \label{eq:fixed stress 3}\\
    \boldsymbol{\alpha}_\mathrm{p}:\boldsymbol{\varepsilon}^{i,j} &= \boldsymbol{\alpha}_\mathrm{p}:\boldsymbol{\varepsilon}^{i,j-1}
    + \boldsymbol{\alpha}_\mathrm{p}:\mathbb{S}_\mathrm{eff}:\boldsymbol{\alpha}_\mathrm{c}\left(p_\mathrm{c}^{i,j} - p_\mathrm{c}^{i,j-1}\right)
    + \boldsymbol{\alpha}_\mathrm{p}:\mathbb{S}_\mathrm{eff}:\boldsymbol{\alpha}_\mathrm{p}\left(p_\mathrm{p}^{i,j} - p_\mathrm{p}^{i,j-1}\right).
    \label{eq:fixed stress 4}
\end{align}

Using Eqs.~\eqref{eq:fixed stress 3} and \eqref{eq:fixed stress 4}, the variations of $\phi_\mathrm{c}$ \eqref{eq:variation_phic} and $\phi_\mathrm{p}$ \eqref{eq:variation_phip} can be written as

\begin{multline}
    \frac{\phi^{i,j}_\mathrm{c} - \phi^{i-1}_\mathrm{c}}{\Delta t}  = \boldsymbol{\alpha}_\mathrm{c}:\frac{\boldsymbol{\varepsilon}^{i,j-1} - \boldsymbol{\varepsilon}^{i-1}}{\Delta t}
    +\left(\boldsymbol{\alpha}_\mathrm{c}:\mathbb{S}_\mathrm{eff}:\boldsymbol{\alpha}_\mathrm{c} + \frac{1}{N_\mathrm{cc}}\right)\frac{p^{i,j}_\mathrm{c} - p^{i-1}_\mathrm{c}}{\Delta t}
    - \boldsymbol{\alpha}_\mathrm{c}:\mathbb{S}_\mathrm{eff}:\boldsymbol{\alpha}_\mathrm{c}\frac{p^{i,j-1}_\mathrm{c} - p^{i-1}_\mathrm{c}}{\Delta t}  \\
    +\left(\boldsymbol{\alpha}_\mathrm{c}:\mathbb{S}_\mathrm{eff}:\boldsymbol{\alpha}_\mathrm{p} + \frac{1}{N_\mathrm{cp}}\right)\frac{p^{i,j}_\mathrm{p} - p^{i-1}_\mathrm{p}}{\Delta t}
    - \boldsymbol{\alpha}_\mathrm{c}:\mathbb{S}_\mathrm{eff}:\boldsymbol{\alpha}_\mathrm{p}\frac{p^{i,j-1}_\mathrm{p} - p^{i-1}_\mathrm{p}}{\Delta t},
    \label{eq:phi_c}
\end{multline}
\begin{multline}
    \frac{\phi^{i,j}_\mathrm{p} - \phi^{i-1}_\mathrm{p}}{\Delta t}  = \boldsymbol{\alpha}_\mathrm{p}:\frac{\boldsymbol{\varepsilon}^{i,j-1} - \boldsymbol{\varepsilon}^{i-1}}{\Delta t}
    +\left(\boldsymbol{\alpha}_\mathrm{p}:\mathbb{S}_\mathrm{eff}:\boldsymbol{\alpha}_\mathrm{p} + \frac{1}{N_\mathrm{pp}}\right)\frac{p^{i,j}_\mathrm{p} - p^{i-1}_\mathrm{p}}{\Delta t}
    - \boldsymbol{\alpha}_\mathrm{p}:\mathbb{S}_\mathrm{eff}:\boldsymbol{\alpha}_\mathrm{p}\frac{p^{i,j-1}_\mathrm{p} - p^{i-1}_\mathrm{p}}{\Delta t}  \\
    +\left(\boldsymbol{\alpha}_\mathrm{p}:\mathbb{S}_\mathrm{eff}:\boldsymbol{\alpha}_\mathrm{c} + \frac{1}{N_\mathrm{pc}}\right)\frac{p^{i,j}_\mathrm{c} - p^{i-1}_\mathrm{c}}{\Delta t}
    - \boldsymbol{\alpha}_\mathrm{p}:\mathbb{S}_\mathrm{eff}:\boldsymbol{\alpha}_\mathrm{c}\frac{p^{i,j-1}_\mathrm{c} - p^{i-1}_\mathrm{c}}{\Delta t}.
    \label{eq:phi_p}
\end{multline}

As can be seen, this system of equations converges to the original system in Eqs.~\eqref{eq:porosity variation M} and \eqref{eq:porosity variation m} when $p^{i,j}_\mathrm{c} \approx p^{i,j-1}_\mathrm{c}$ and $p^{i,j}_\mathrm{p} \approx p^{i,j-1}_\mathrm{p}$.

Substituting Eqs.~\eqref{eq:phi_c} and \eqref{eq:phi_p} into Eq.~\eqref{eq:mass_balance}, we have the temporal discretized equation as
\begin{multline}
    \boldsymbol{\alpha}_\mathrm{c}:\frac{\boldsymbol{\varepsilon}^{i,j-1} - \boldsymbol{\varepsilon}^{i-1}}{\Delta t}
    +\left(\boldsymbol{\alpha}_\mathrm{c}:\mathbb{S}_\mathrm{eff}:\boldsymbol{\alpha}_\mathrm{c} + \frac{1}{N_\mathrm{cc}}\right)\frac{p^{i,j}_\mathrm{c} - p^{i-1}_\mathrm{c}}{\Delta t}
    - \boldsymbol{\alpha}_\mathrm{c}:\mathbb{S}_\mathrm{eff}:\boldsymbol{\alpha}_\mathrm{c}\frac{p^{i,j-1}_\mathrm{c} - p^{i-1}_\mathrm{c}}{\Delta t} \\
    +\left(\boldsymbol{\alpha}_\mathrm{c}:\mathbb{S}_\mathrm{eff}:\boldsymbol{\alpha}_\mathrm{p} + \frac{1}{N_\mathrm{cp}}\right)\frac{p^{i,j}_\mathrm{p} - p^{i-1}_\mathrm{p}}{\Delta t}
    - \boldsymbol{\alpha}_\mathrm{c}:\mathbb{S}_\mathrm{eff}:\boldsymbol{\alpha}_\mathrm{p}\frac{p^{i,j-1}_\mathrm{p} - p^{i-1}_\mathrm{p}}{\Delta t}
    -\nabla \cdot\frac{K_\mathrm{c}}{\mu_f}\left( \nabla p^{i,j}_\mathrm{c} - \rho_f \mathbf{g}  \right) \\ -\frac{ K_\mathrm{m}}{\mu} \frac{V}{h} \frac{2(p^{i,j}_\mathrm{p} - p^{i,j}_\mathrm{c})}{h/2} = 0,
    \label{eq:phi_c_temporal}
\end{multline}
\begin{multline}
    \boldsymbol{\alpha}_\mathrm{p}:\frac{\boldsymbol{\varepsilon}^{i,j-1} - \boldsymbol{\varepsilon}^{i-1}}{\Delta t}
    +\left(\boldsymbol{\alpha}_\mathrm{p}:\mathbb{S}_\mathrm{eff}:\boldsymbol{\alpha}_\mathrm{p} + \frac{1}{N_\mathrm{pp}}\right)\frac{p^{i,j}_\mathrm{p} - p^{i-1}_\mathrm{p}}{\Delta t}
    - \boldsymbol{\alpha}_\mathrm{p}:\mathbb{S}_\mathrm{eff}:\boldsymbol{\alpha}_\mathrm{p}\frac{p^{i,j-1}_\mathrm{p} - p^{i-1}_\mathrm{p}}{\Delta t} \\
    +\left(\boldsymbol{\alpha}_\mathrm{p}:\mathbb{S}_\mathrm{eff}:\boldsymbol{\alpha}_\mathrm{c} + \frac{1}{N_\mathrm{pc}}\right)\frac{p^{i,j}_\mathrm{c} - p^{i-1}_\mathrm{c}}{\Delta t}
    - \boldsymbol{\alpha}_\mathrm{p}:\mathbb{S}_\mathrm{eff}:\boldsymbol{\alpha}_\mathrm{c}\frac{p^{i,j-1}_\mathrm{c} - p^{i-1}_\mathrm{c}}{\Delta t}
    -\nabla \cdot\frac{K_\mathrm{p}}{\mu_f}\left( \nabla p^{i,j}_\mathrm{p} - \rho_f \mathbf{g}  \right) \\ -\frac{ K_\mathrm{m}}{\mu} \frac{V}{h} \frac{2(p^{i,j}_\mathrm{c} - p^{i,j}_\mathrm{p})}{h/2} = 0.
    \label{eq:phi_p_temporal}
\end{multline}

\subsection{Dynamic update of the solution domain of fracture flow}
\label{sec:dynamic_update}

In the double-porosity framework, the mesocrack pressure $p_\mathrm{c}$ and the micropore pressure $p_\mathrm{p}$ coexist throughout the domain $\Omega$.
However, the fracture flow, Eq.~\eqref{eq:phi_c_temporal}, is physically meaningful only within the fractured zone, where the phase field has evolved sufficiently to open a macroscopic fracture.
In the intact region, no fracture exists, thereby solving Eq.~\eqref{eq:phi_c_temporal} would introduce spurious fluid transport. 
To address this, we introduce a dynamic phase-field--dependent indicator function $\chi: [0,1]\to\{0,1\}$ to ensure that the fracture flow domain keep pace with the propagating crack without any user-defined geometric prescription of the fracture path:
\begin{equation}
    \chi(v) =
    \begin{cases}
        1 & \text{if } v \leq v_\mathrm{cr}, \\
        0 & \text{if } v > v_\mathrm{cr},
    \end{cases}
    \label{eq:chi}
\end{equation}
where $v_\mathrm{cr} \in (0,1)$ is a prescribed threshold. The condition $v \leq v_\mathrm{cr}$ identifies the fractured zone $\Omega_\Gamma = \{x\in\Omega : v(x) \leq v_\mathrm{cr}\}$, while $\Omega \setminus \Omega_\Gamma$ denotes the intact region.
The fracture flow Eq.~\eqref{eq:phi_c_temporal} is then replaced by its indicator-weighted counterpart:
\begin{equation}
    \chi(v)\,\mathcal{L}_\mathrm{c}(p_\mathrm{c}, p_\mathrm{p}, \boldsymbol{\varepsilon}) = 0 \quad \text{in } \Omega,
    \label{eq:phi_c_modified} 
\end{equation}
where $\mathcal{L}_\mathrm{c}$ denotes the left-hand side of Eq.~\eqref{eq:phi_c_temporal}.
Inside the fractured zone ($\chi = 1$), Eq.~\eqref{eq:phi_c_modified} reduces to the original fracture flow equation, so $p_\mathrm{c}$ is governed by the full poromechanical coupling including Darcy flow and inter-scale fluid exchange.
Outside the fractured zone ($\chi = 0$), Eq.~\eqref{eq:phi_c_modified} is trivially satisfied, and $p_\mathrm{c}$ loses its governing equation.
In this region, the mesocrack and micropore networks are in local pressure equilibrium, so we set
\begin{equation}
    p_\mathrm{c} = p_\mathrm{p} \quad \text{in } \Omega \setminus \Omega_\Gamma.
    \label{eq:pc_intact}
\end{equation}
Note that this condition holds only for eliminating the inter-scale flux term in Eq.~\eqref{eq:phi_p_temporal} outside the fracture ($\Omega\setminus\Omega_\Gamma$) , recovering the behavior of a single-porosity medium in the intact region.



\subsection{Variational inequality solution for fluid lag}
\label{sec:fluid_lag}

In hydraulic fracturing, the advancing fracturing tip may outpace the fluid front, leaving a near-tip lag zone where the fracture is mechanically open but devoid of pressurized fluid, known as fluid lag (Fig.~\ref{fig:parametric_space}a).
Within this lag zone, the fluid pressure reduces to zero (or gas pressure), since fluid cannot sustain tensile stresses (cavitation). 
Without an explicit non-negativity constraint, the numerical solution of Eqs.~\eqref{eq:phi_c_temporal} and~\eqref{eq:phi_p_temporal} may yield negative pressures in the lag zone, which is physically inadmissible.
To preclude such spurious states, the flow sub-problem is reformulated as a variational inequality, enforcing $p_\mathrm{c} \geq 0$ and $p_\mathrm{p} \geq 0$ throughout $\Omega$.

Let $\mathcal{V} = H^1(\Omega)$.
We introduce the admissible convex cones:
\begin{equation}
    \mathcal{A}_\mathrm{c} = \{ q \in \mathcal{V} : q \geq 0 \;\text{a.e.\ in } \Omega \}, \qquad
    \mathcal{A}_\mathrm{p} = \{ q \in \mathcal{V} : q \geq 0 \;\text{a.e.\ in } \Omega \},
    \label{eq:admissible_sets}
\end{equation}
which encode the physical requirement that both pressure fields remain non-negative.
The variational inequality problem then reads: find $(p_\mathrm{c}, p_\mathrm{p}) \in \mathcal{A}_\mathrm{c} \times \mathcal{A}_\mathrm{p}$ such that
\begin{align}
    \int_\Omega \chi(v)\mathcal{L}_\mathrm{c}(p_\mathrm{c}, p_\mathrm{p}, \boldsymbol{\varepsilon})\,(q_\mathrm{c} - p_\mathrm{c})\,\mathrm{d}\Omega &\geq 0 \quad \forall\, q_\mathrm{c} \in \mathcal{A}_\mathrm{c}, \label{eq:vi_c} \\
    \int_\Omega \mathcal{L}_\mathrm{p}(p_\mathrm{c}, p_\mathrm{p}, \boldsymbol{\varepsilon})\,(q_\mathrm{p} - p_\mathrm{p})\,\mathrm{d}\Omega &\geq 0 \quad \forall\, q_\mathrm{p} \in \mathcal{A}_\mathrm{p}, \label{eq:vi_p}
\end{align}
where $\mathcal{L}_\mathrm{p}$ is the  left-hand side of Eq.~\eqref{eq:phi_p_temporal}.

The Karush--Kuhn--Tucker (KKT) conditions associated with Eqs.~\eqref{eq:vi_c} and \eqref{eq:vi_p} yield the pointwise complementarity system
\begin{align}
    &p_\mathrm{c} \geq 0, \quad \chi(v)\,\mathcal{L}_\mathrm{c} \geq 0, \quad p_\mathrm{c}\,[\chi(v)\,\mathcal{L}_\mathrm{c}] = 0 \quad \text{in } \Omega, \label{eq:kkt_c} \\
    &p_\mathrm{p} \geq 0, \quad \mathcal{L}_\mathrm{p} \geq 0, \quad p_\mathrm{p}\,\mathcal{L}_\mathrm{p} = 0 \quad \text{in } \Omega. \label{eq:kkt_p}
\end{align}
These conditions naturally partition the domain into two zones.
In the fluid-filled zone, $p > 0$ and the flow equation is satisfied as an equality ($\mathcal{L} = 0$).
In the lag zone, $p = 0$ and the residual satisfies $\mathcal{L} \geq 0$. 
The problems, Eqs.~\eqref{eq:kkt_c} and~\eqref{eq:kkt_p} are solved numerically by using the reduced space active set solvers of PETSc~\citep{petsc-user-ref} for variational inequalities based on Newton’s method.



\section{Numerical verification} \label{sec:examples}
In this section, we focus on the plane-strain hydraulic fracturing problem (KGD problem) and present three numerical examples to verify the proposed model across different regimes, as in~\cite{garagash2006plane}. 
The first example focuses on toughness-dominated hydraulic fracturing (K-vertex)\footnote{The toughness-dominated regime is considered ``the trivial solution of a uniformly pressurized fracture with zero lag''~\citep{lecampion2007implicit} compared to the other regimes.} without leak-off (i.e., $\mathcal{S}=0$) to demonstrate that the proposed model recovers the solutions achieved by standard single-continuum phase-field hydraulic fracturing models.
The second example illustrates the capability of the proposed framework to simulate viscosity-dominated hydraulic fracture (M-vertex), a regime that existing phase-field models struggle to capture. 
The final example demonstrates that the proposed model successfully captures fluid lag (O-vertex), a phenomenon that, to the best of our knowledge, has not yet been studied by any existing phase-field models.

\subsection{Closed-form expression}

We first present the closed-form expressions. By using viscosity scaling and considering the far-field confining pressure $\sigma_\mathrm{N}$, the half-length of the fracture $L(t)$, the half-length of fluid-filled region $L_{f}(t)$, the net pressure $p(x,t)-\sigma_\mathrm{N}$ and the fracture opening $w(x,t)$ can be expressed in terms of the dimensionless terms as follows~\citep{garagash2006propagation}:
\begin{align}
L(t) &= L_\mathrm{m}(t) \gamma_\mathrm{m}(t), \label{eq:Lt} \\
L_{f}(t) &= L_\mathrm{m}(t) \gamma_{f\mathrm{m}}(t), \\
p(x,t) - \sigma_\mathrm{N} &= \epsilon_\mathrm{m}(t) E^\prime \Pi_\mathrm{m}(\xi,t), \label{eq:pt}\\
w(x,t) &= \epsilon_\mathrm{m}(t) L_\mathrm{m}(t) \Omega_\mathrm{m}(\xi,t). \label{eq:wt}
\end{align}
where the dimensionless spatial coordinate is $\xi = x/L(t)$ and the viscosity scaling is defined as 
\begin{equation}
L_\mathrm{m} = \left( \frac{{E^\prime}^{1/4} Q^{3/4} t}{{\mu^\prime}^{1/4}} \right)^{2/3},\,  
\epsilon_\mathrm{m} = \left( \frac{\mu^\prime}{E^\prime t} \right)^{1/3},\, 
\mathcal{K}_\mathrm{m} = \frac{K^\prime}{{E^\prime}^{3/4} Q^{1/4} \mu^{1/4}},\, 
\mathcal{T}_\mathrm{m} = \frac{\sigma_\mathrm{N} - p_\mathrm{tip}}{E^\prime} \left( \frac{E^\prime t}{\mu^\prime} \right)^{1/3} \label{eq:K_m}
\end{equation} 
where the parameters $K^\prime$, $E^\prime$ and $\mu^\prime$ are defined as
\begin{equation}
     K^\prime = 4\sqrt{\frac{2}{\pi}}K_\mathrm{IC}, \, K_\mathrm{IC}=\sqrt{\frac{EG_c}{1-\nu^2}}, \, E^\prime = \frac{E}{1-\nu^2}, \, \mu^\prime = 12 \mu, 
\end{equation}
and $\mu$, $Q$ and $G_c$ are the fluid dynamic viscosity, injection flow rate, and critical energy release rate, respectively.
For simplicity, we consider zero vapor pressure, i.e., $p_\mathrm{tip} = 0$ in this study.

The dimensionless terms $\gamma_\mathrm{m}(t)$, $\gamma_{f\mathrm{m}}(t)$, $\Pi_\mathrm{m}(\xi,t)$, and $\Omega_\mathrm{m}(\xi,t)$ in Eqs.~\eqref{eq:Lt}--\eqref{eq:wt} depend on the propagation regime. 
These self-similar asymptotic solutions for the plane-strain Kristianovich--Geertsma--de Klerk (KGD) problem at the M-vertex~\citep{garagash2005plane}, the K-vertex~\citep{garagash2006plane}, and the early-time O-vertex~\citep{garagash2006propagation} not only have served as canonical benchmarks for numerical models but also have advanced our insight into hydraulic fracture mechanics.

\subsection{Numerical verification cases}

\begin{figure}[!ht]
    \centering    \includegraphics[width=0.7\linewidth]{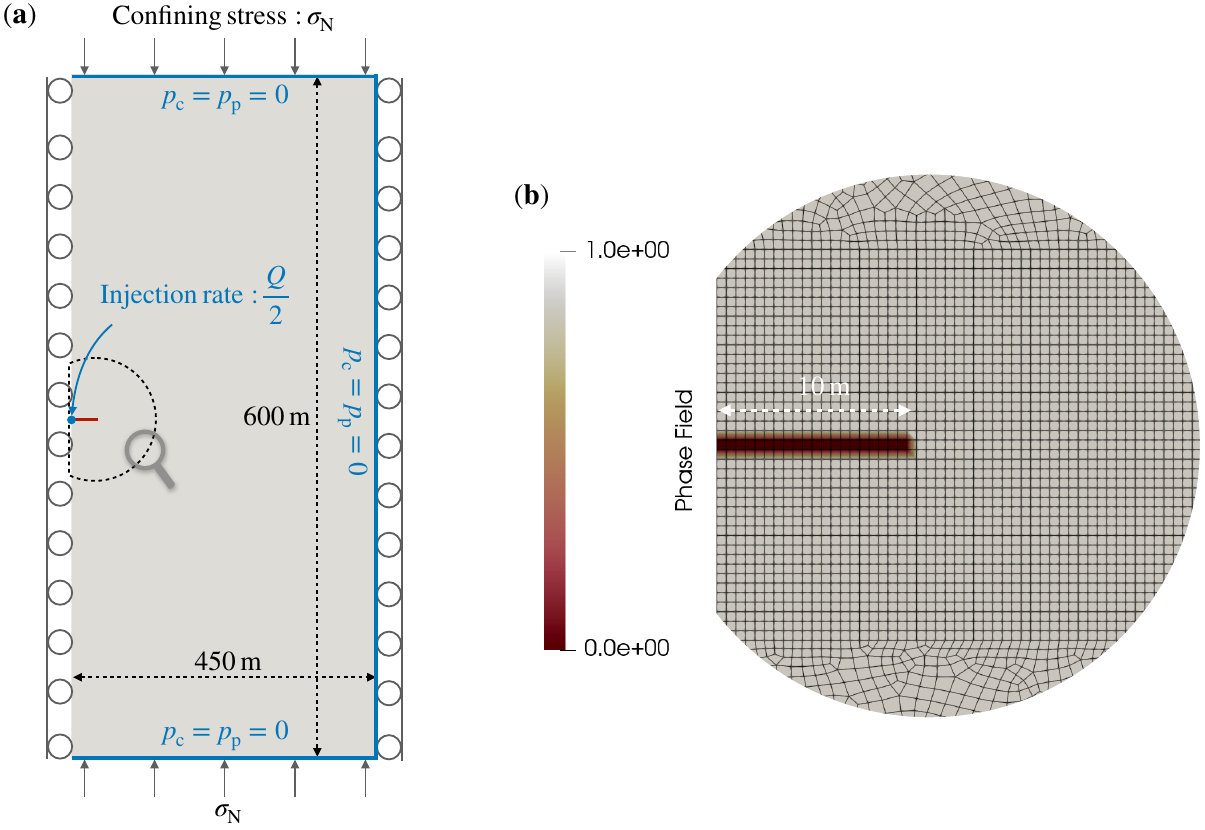}
    \caption{The geometry and boundary conditions of the plane-strain hydraulic fracture problem (a), and the initial phase-field of the discretized model (b).}
    \label{fig:model_setting}
\end{figure}

The geometry, boundary conditions, mesh, and initial settings of the computational model are illustrated in Fig.~\ref{fig:model_setting}a. The domain size is 450 m $\times$ 600 m. The left and right edges are constrained in the normal direction. 
The top and bottom edges of the model are subjected to confining pressure $\sigma_\mathrm{N}$. 
All the boundaries are drained except the left one.  
Fluid is injected at a constant rate of $Q/2$ from the left side of the initial fracture which is specified by an initial phase-field value. 
The length of the initial fracture is 10 m. The elements are refined along the fracture propagation path, with a spacing of $h = 0.5$ m (Fig.~\ref{fig:model_setting}b). Unless otherwise specified, the material properties for the toughness-dominated and viscosity-dominated numerical cases are listed in Table~\ref{tab:parameters}.

\begin{table}[!h]
\centering
\caption{Material parameters for the numerical examples.}
\label{tab:parameters}
\begin{tabular}{llccc}
\toprule
Parameter & Symbol & Unit & Toughness-dominated & Viscosity-dominated \\
\midrule
Young's modulus              & $E$                    & GPa                          & 17                    & 17                    \\
Poisson's ratio              & $\nu$                  & ---                          & 0.2                   & 0.2   \\
Phase-field length scale     & $\ell$                 & m                            & 1.0                   & 1.0                 \\
Injection rate               & $Q$         & \si{m^2 \per s}              & $2\times10^{-3}$      & $2\times10^{-3}$      \\
Matrix permeability       & $K_\mathrm{p}$         & \si{m^2}                     & $10^{-19}$            & $10^{-19}$            \\
Biot coefficient of the matrix      & $\alpha_\mathrm{m}$    & ---                          & 0                     & 0                     \\
Matrix porosity              & $\phi_\mathrm{m}$      & ---                          & 0                     & 0                     \\
\midrule
Fluid viscosity              & $\mu$                  & \si{Pa \cdot s}              & $10^{-8}$             & $10^{-2}$             \\
Critical energy release rate & $G_\mathrm{c}$         & \si{Pa \cdot m}              & 120                   & 60                    \\
Confining stress             & $\sigma_\mathrm{N}$    & \si{MPa}                     & 0                     & 0.8                   \\
\bottomrule
\end{tabular}
\end{table}

\subsection{Toughness-dominated regime}

\cite{garagash2005plane} introduced a dimensionless viscosity $\mathcal{K}_\mathrm{m}$ (defined in Eq.~\eqref{eq:K_m}$_3$) to distinguish different hydraulic fracture regimes. 
Note that the $\mathcal{K}_\mathrm{m}$ value in this study is computed using the effective critical energy release rate $G_\mathrm{c}^{\mathrm{eff}} = G_\mathrm{c}(1 + 3h_\mathrm{e}/8 \ell)$ with element size $h = \SI{0.5}{m}$ and phase-field length scale $\ell = \SI{1.0}{m}$ to account for the discretization error in the phase-field approximation of the fracture energy~\citep{Bourdin2008, Yoshioka2020}.

The toughness- and viscosity--dominated regimes can then be classified according to
\begin{align}
   \text{toughness-dominated regime:}&  \, \mathcal{K}_\mathrm{m}  > 4.13, & \\
   \text{viscous-dominated regime:} & \, \mathcal{K}_\mathrm{m} < 0.7, \label{eq:km<0.7}&
\end{align}
under which the self-similar analytical solutions for the fracture length, aperture, and pressure can be derived~\citep{garagash2005plane}.

\begin{figure}[!h]
    \centering
\includegraphics[width=1.0\linewidth]{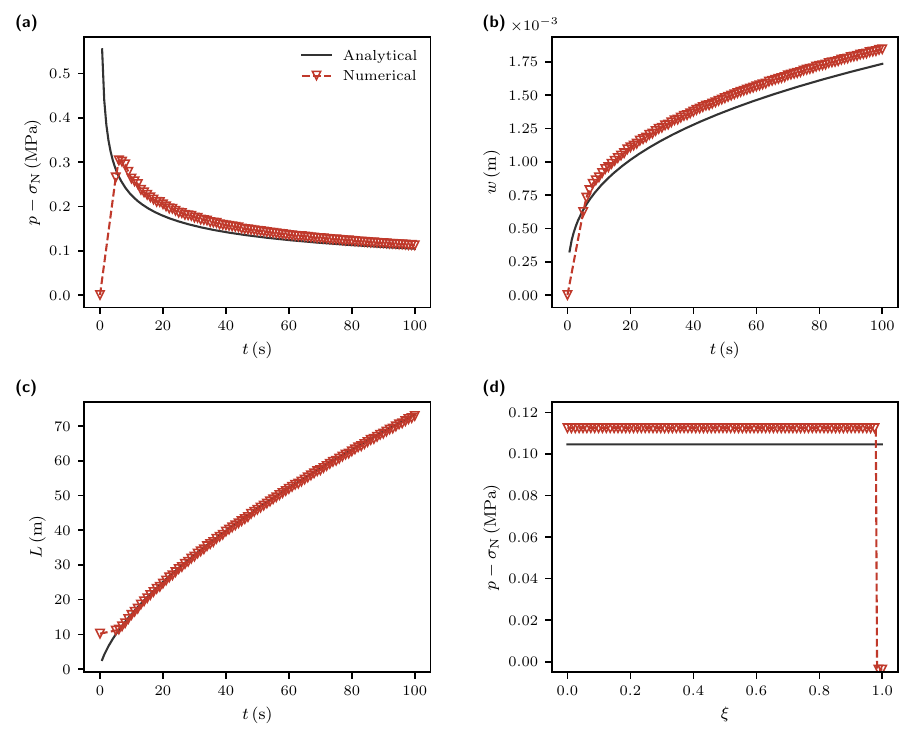}
    \caption{The analytical and numerical comparisons of the pressure at the injection (a), fracture aperture at the injection point (b), fracture length (c), and pressure profile along the fracture at time 100 s (d) in the toughness-dominated regime. }
    \label{fig:KGD_toughness_result}
\end{figure}

Using the parameters in Table~\ref{tab:parameters} for the toughness-dominated case, we have $\mathcal{K}_\mathrm{m} = 142.5$, which is located in the K-vertex. In this simulation, the initial time increment is set to \SI{5}{s}, followed by uniform increments of \SI{1}{s} for the remainder of the analysis. This time setting allows fluid to rapidly fill the pre-existing fracture, thereby capturing subsequent fracture propagation more accurately.

Fig.~\ref{fig:KGD_toughness_result} shows comparisons between the numerical and analytical solutions for the pressure at the injection point, fracture aperture at the injection point, fracture length, and pressure profile along the fracture at time 100. 
The numerical results agree well with the analytical solutions, demonstrating that the proposed model can capture the behavior of toughness-dominated hydraulic fracturing. 
The small discrepancies in the pressure and aperture at the injection point is attributed to a slight delay of the initiation of the hydraulic fracture as present in the analytical solution. 
The fracture length matches almost perfectly with the analytical solution, and the pressure distribution along the fracture is uniform, indicating that the overall fracture propagation driven by pressurization is well captured by the numerical model.

\subsection{Viscosity-dominated regime}

For this case, $\mathcal{K}_\mathrm{m} = 0.593$, placing it in the viscosity-dominated regime according to Eq.~\eqref{eq:km<0.7}. 
Unlike the viscosity-dominated solution presented in~\cite{garagash2005plane}, which assumes no fluid lag, our numerical model does not prohibit the fluid lag development. 
Thus, for a fair comparison with the analytical solution, we applied a confining stress of \SI{0.8}{MPa} to the top and bottom boundaries to prevent the fracture from propagating too rapidly, which minimizes the fluid lag.

\begin{figure}[!ht]
    \centering
\includegraphics[width=1.0\linewidth]{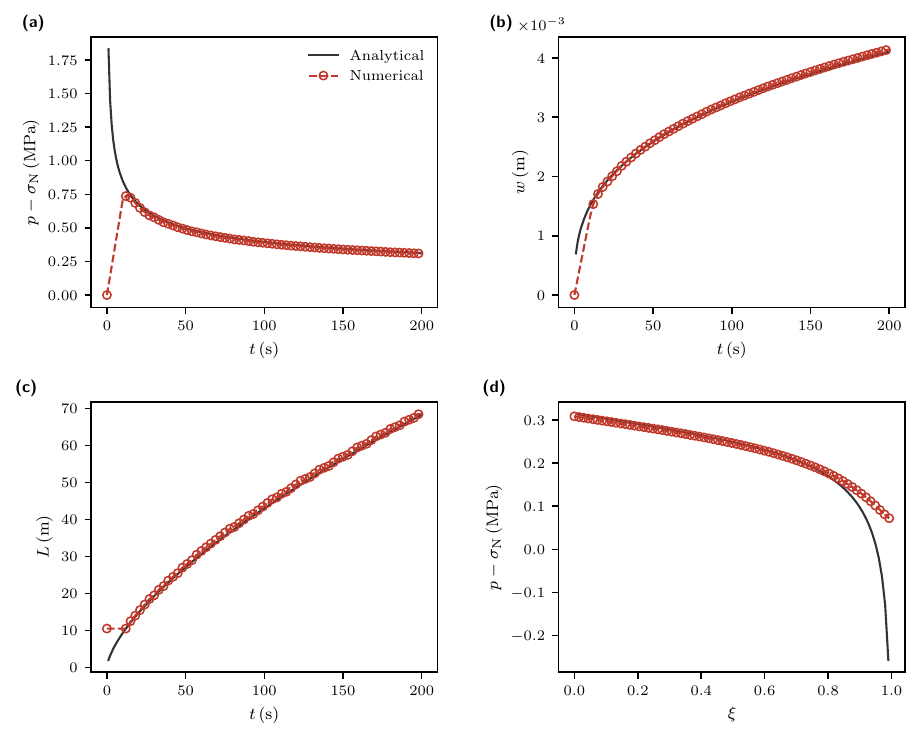}
    \caption{The analytical and numerical comparisons of the pressure at the injection (a), fracture aperture at the injection point (b), fracture length (c), and pressure profile along the fracture at time 100 s (d) in the viscosity-dominated domain.}
    \label{fig:KGD_viscous_result}
\end{figure}

The simulation employed an initial time increment of \SI{10}{s}, followed by uniform increments of \SI{1}{s}. Fig.~\ref{fig:KGD_viscous_result} compares the numerical and analytical solutions for the injection pressure, fracture aperture at the injection point, fracture length evolution, and the spatial pressure profile along the fracture at 100 s. 
The close agreement demonstrates that the proposed model captures the behavior of viscosity-dominated hydraulic fracturing. 
Notably, the breakdown pressure matches the analytical solution more closely than in the toughness-domianted case (Fig.~\ref{fig:KGD_viscous_result}a). 
This is because the energy dissipation in this viscosity-dominated regime is dominated by fluid viscous dissipation, which is less sensitive to the phase-field regularization of the fracture tip. 
Furthermore, the pressure distribution along the fracture is non-uniform, the highest at the injection point with a steady decrease towards the fracture tip due to the viscous resistance (Fig.~\ref{fig:KGD_viscous_result}d).
This pressure gradient is absent in the toughness-dominated case, where the primary energy dissipation is fracture surface creation rather than viscous fluid flow. 

The pressure distribution along the fracture ($\xi$\footnote{The horizontal axis $\xi = x/L \in [0,1]$ is the dimensionless coordinate along the fracture half-length, with $\xi = 0$ at the injection point and $\xi = 1$ at the fracture tip.}) agrees well with the analytical solution (Fig.~\ref{fig:KGD_viscous_result}d), except near the fracture tip ($\xi > 0.9$).
The analytical solution yields a negative net pressure near the fracture tip because it assumes the absence of a fluid lag zone; consequently, it is only applicable in the region away from the fracture tip~\citep{garagash2005plane}. 
In contrast, the numerical model inherently accommodates the lag effect, and therefore prevents the unphysical negative net pressure near the fracture tip, which explains the deviation from the analytical solution.

\begin{figure}[!htp]
    \centering
\includegraphics[width=0.6\linewidth]{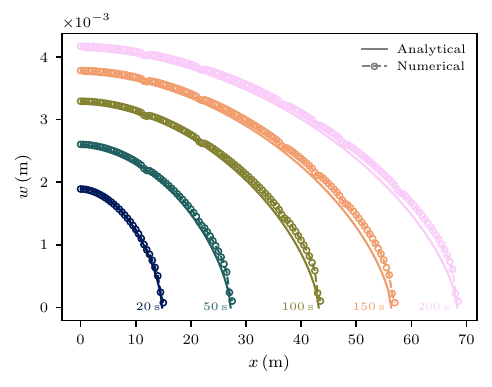}
    \caption{The fracture aperture profiles along the fracture at different time frames from 20 s to 200 s in the viscosity-dominated regime without fluid lag.}
    \label{fig:w_profile}
\end{figure}

In addition to the comparing the fracture aperture evolution at the injection point, the spatial aperture profiles are compared with the analytical solutions from 20 s to 200 s in Fig.~\ref{fig:w_profile}. 
Our model closely captures increasing evolution of the aperture profiles predicted by the analytical solutions of viscosity-dominated hydraulic fracturing.
These spatial and temporal comparisons confirm that both fluid flow and fracture propagation are well captured by the proposed model.

\subsection{Hydraulic fracture with fluid lag}
This example examines the behavior of hydraulic fractures with fluid lag.
A fluid front may lag behind the fracture tip when the fluid viscosity is high and the confining stresses are low, such as in near-surface hydraulic fracturing~\citep{bunger2013comparison} or magma dykes~\citep{rubin1995propagation}.  
The parameters for this simulation are the same as those in the viscosity-dominated case in Table~\ref{tab:parameters}, except that no confining stress is applied to allow fluid lag development ($\mathcal{T}_\mathrm{m} \approx 0$).
To investigate how the size of the fluid lag zone varies with fracture toughness, we consider four values of the critical energy release rate, $G_\mathrm{c} \in \{60, 42, 28, 17\}$ \si{Pa \cdot m}.
The corresponding dimensionless toughness $\mathcal{K}_\mathrm{m}$ are listed in Table~\ref{tab:lag_parameters}.
These four cases are chosen to match the $\mathcal{K}_\mathrm{m}$ values presented by~\cite{garagash2006propagation} (Table 2 in their manuscript) in order to evaluate the numerical model against the analytical solutions and explore the influence of toughness on the extent of the fluid lag zone.

\begin{table}[!h]
\centering
\caption{Critical energy release rate and corresponding dimensionless toughness for the four fluid-lag cases ($h_e = \SI{0.5}{m}$, $\ell = \SI{1.0}{m}$). All other parameters are identical to those of the viscosity-dominated case in Table~\ref{tab:parameters}.}
\label{tab:lag_parameters}
\begin{tabular}{cccc}
\toprule
Case & $G_\mathrm{c}$ (\si{Pa \cdot m}) & $G_\mathrm{c}^{\mathrm{eff}}$ (\si{Pa \cdot m}) & $\mathcal{K}_\mathrm{m}$ \\
\midrule
1 & 60 & 71.25 & 0.593 \\
2 & 42 & 49.88 & 0.496 \\
3 & 28 & 33.25 & 0.405 \\
4 & 17 & 20.19 & 0.316 \\
\bottomrule
\end{tabular}
\end{table}

\begin{figure}[!h]
    \centering
\includegraphics[width=1.0\linewidth]{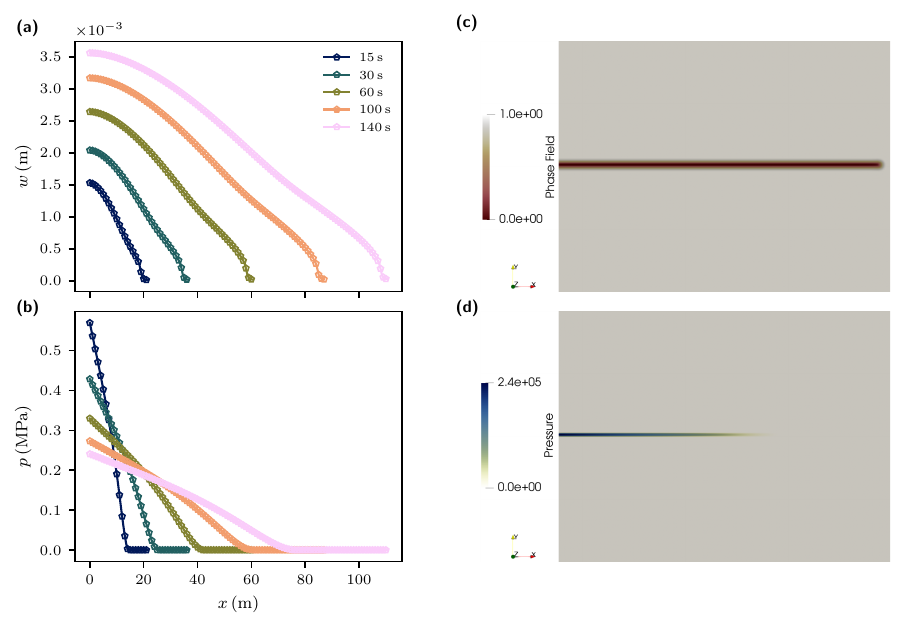}
    \caption{The fracture aperture profile (a) and pressure profile (b) along the hydraulic fractures at different times for Case 1 as listed in Table~\ref{tab:lag_parameters}, and the phase-field (b) and pressure (c) distribution at 140 s.}
    \label{fig:fluid_lag_profiles}
\end{figure}

The initial time increment is \SI{10}{s}, followed by a uniform increment of \SI{1}{s} for the remainder of the analysis. 
For Case 1, Fig.~\ref{fig:fluid_lag_profiles} presents the evolution of the fracture aperture and the fluid pressure distribution over the time interval $[15, 140]$~s (Fig.~\ref{fig:fluid_lag_profiles}a and b), together with the phase-field and pressure contours at the final time of 140~s (Fig.~\ref{fig:fluid_lag_profiles}c and d). 
The aperture profiles in Fig.~\ref{fig:fluid_lag_profiles}a exhibit the smooth, concave shape characteristic of viscosity-dominated fracture growth. The opening is greatest at the injection point and decreases monotonically toward the fracture tip.
The aperture at $x = 0$ increases from approximately $1.5 \times 10^{-3}$~m at 15~s to $3.5 \times 10^{-3}$~m at 140~s, and the fracture half-length grows from roughly 15~m to 110~m over the same period.

The pressure profiles in Fig.~\ref{fig:fluid_lag_profiles}b reveal two distinct zones along the fracture: a pressurized fluid-filled zone and a near-tip lag zone where the fluid pressure is zero despite the fracture being mechanically open ($w > 0$).
The relative length of the lag zone compared with the fracture length remains the same, illustrating the self-similar nature of the analytical solution~\citep{garagash2006propagation}.
Fig.~\ref{fig:fluid_lag_profiles}c shows the phase-field distribution at 140~s, where the smeared crack band is well resolved without any numerical instability.
Fig.~\ref{fig:fluid_lag_profiles}d presents the corresponding pressure field $p_\mathrm{c}$ in fracture at 140~s, displaying a pronounced gradient from the injection point towards the fracture tip.
The fluid lag can be clearly observed by comparing Fig.~\ref{fig:fluid_lag_profiles}c against Fig.~\ref{fig:fluid_lag_profiles}d as the near-zero pressure region adjacent to the crack tip, corroborating the one-dimensional profiles in Fig.~\ref{fig:fluid_lag_profiles}b.
The sharp transition from a pressurized fracture to an unpressurized, open fracture is captured by the proposed model, confirming that the dynamic domain update (Section~\ref{sec:dynamic_update}) correctly confines the fracture flow equation to the fractured zone ($\Omega_\Gamma$) while the variational inequality formulation (Section~\ref{sec:fluid_lag}) enforces zero pressure in the lag zone without explicitly prescribing the fluid front.


\begin{figure}[!htb]
    \centering
\includegraphics[width=0.87\linewidth]{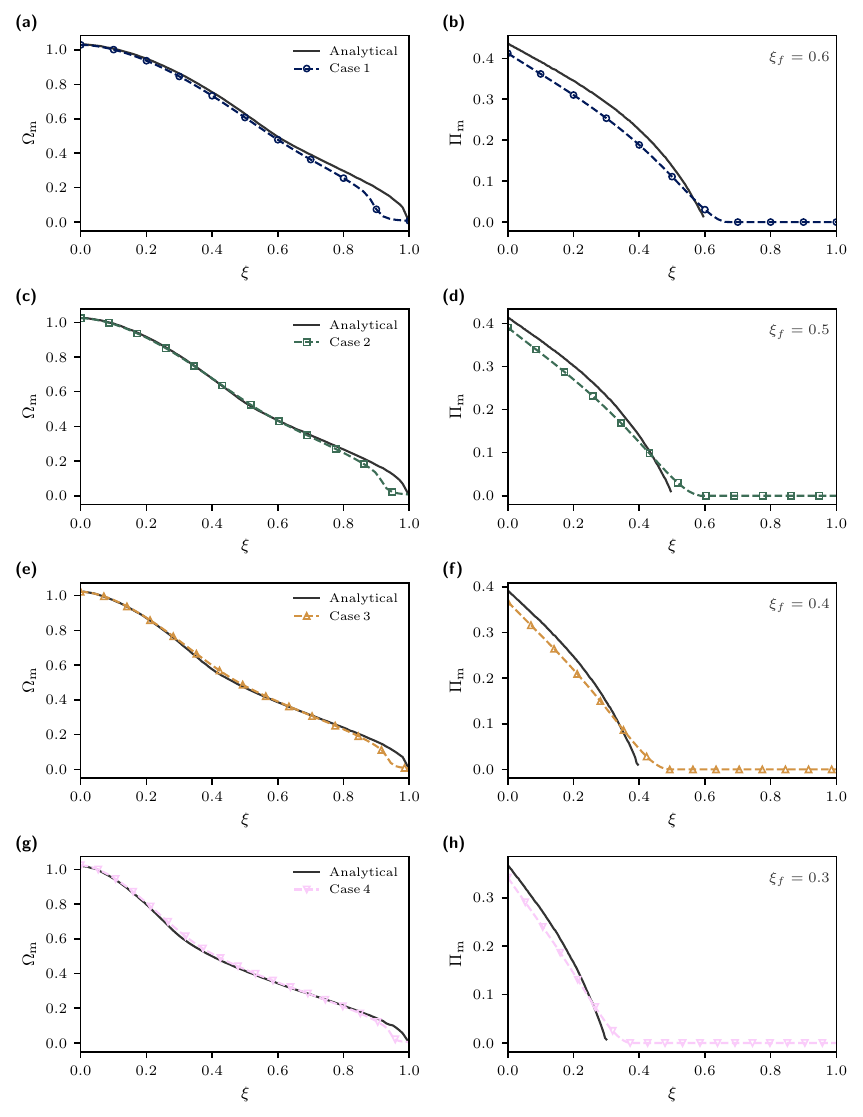}
    \caption{The comparisons of the scaled pressure (left column) and fracture aperture (right column) along the fracture between the numerical at $t=18$~s and analytical solutions in the 4 cases, where the analytical solutions for 4 different $\xi_f$ in each row ranging from 0.3 to 0.6 are extracted from Fig. 5 and Fig. 6 of ~\cite{garagash2006plane}.}
    \label{fig:Omega_Pi_comparison}
\end{figure}

To assess the accuracy of the proposed model, Fig.~\ref{fig:Omega_Pi_comparison} compares the numerically computed aperture and pressure profiles for all four cases against the self-similar analytical solutions of~\cite{garagash2006plane} in dimensionless form. 
The vertical axes $\Omega_\mathrm{m}$ and $\Pi_\mathrm{m}$ denote the dimensionless aperture and pressure, normalized by the characteristic scales of the viscosity-dominated (M-vertex) solution. 
The numerical results of $\Omega_\mathrm{m}$ and $\Pi_\mathrm{m}$ are post-processed using Eqs.~\eqref{eq:wt} and \eqref{eq:pt}. 
The time step for this post-processing is at 18 s, which is selected to match the analytical solutions of $\Omega_\mathrm{m}$ as closely as possible. 
The analytical family corresponds to dimensionless fluid-front position $\xi_f \in \{0.3, 0.4, 0.5, 0.6\}$, where $\xi_f = L_f/L$ is the ratio of the fluid front to the crack tip location, where $\xi_f = 1$ corresponds to a fully fluid-filled fracture (zero lag), and decreasing $\xi_f$ corresponds to a progressively larger lag zone.

Figs.~\ref{fig:Omega_Pi_comparison}a, c, e, and g show that the numerically calculated aperture profiles for Cases 1--4 fall consistently within the analytical family, ranging from the $\xi_f = 0.6$ (Fig.~\ref{fig:Omega_Pi_comparison}a) to $\xi_f = 0.3$ (Fig.~\ref{fig:Omega_Pi_comparison}g). 
All aperture profiles exhibit the expected smooth decay from a near-unit value at the injection point to zero at the crack tip. The size of the lag zone increases with decreasing $\mathcal{K}_\mathrm{m}$, in agreement with the analytical predictions.
Figs.~\ref{fig:Omega_Pi_comparison}b, d, f, and h reinforce these observations through the pressure profile comparisons. 
The trend is physically consistent, i.e., lower fracture toughness allows the crack tip to advance more readily relative to the fluid front, enlarging the lag zone.

Although the current model can capture hydraulic fracturing with fluid lag, it underestimates the size of the fluid lag zone, as seen in Figs.~\ref{fig:Omega_Pi_comparison}b, d, f, and h, where the simulated fluid front ($\xi_f$) is more advanced than the corresponding analytical solution, and the pressure profile along the fracture is slightly more diffuse. 
This discrepancy may be attributed to non-strict enforcement of the mass conservation condition, such as zero flux from the pressurized zone to the lag zone. 
Additionally, numerical diffusion may also contribute to this discrepancy. 
Future work could incorporate advanced cavity detection methods, such as Elrod–Adams model~\citep{schultz2025new}, or adaptive mesh refinement near the fracture tip to mitigate these effects and improve the accuracy of lag zone simulation.

\section{Conclusions} \label{sec:conclusion}

This paper proposes a variationally consistent dual-continuum phase-field model for hydraulic fracturing in two-scale (double-porosity) porous media, accommodating two independent fluid pressures $p_\mathrm{c}$ and $p_\mathrm{p}$ whose governing parameters are derived rationally from microporomechanics.
The fixed-stress split scheme is extended to solve this two-pressure system, while a variational inequality combined with a dynamic domain update enforces non-negative pressures in the fluid lag zone. 
This formulation effectively describes fluid flow in the fracture zone and accurately reproduces the fluid lag behavior without explicit front tracking.
The proposed model was verified against three plane-strain (KGD) solutions across the toughness-dominated (K-vertex), viscosity-dominated (M-vertex), and early-time (O-vertex) regimes. 
The results demonstrate that the model agrees well with the analytical solutions, correctly captures the parametric dependence of the lag zone on fracture toughness. 
Furthermore, accounting for the pressure discontinuity across the fracture and the porous media prevents phase-field profile instability under the viscosity-dominated regime (M-vertex).  
This double-porosity framework provides a promising tool for understanding and predicting hydraulic fracturing under realistic stress conditions and varying fluid properties, where the viscous dissipation and/or the fluid lag effect may be significant.

For fluid lag cases, the tendency to overestimate the pressurized zone size suggests that incorporating a more rigorous cavitation detection method is a promising direction for future improvement.
Beyond these numerical refinements, two physically motivated extensions are of particular interest.
First, activating fluid leak-off between the mesocrack network and the micropore space will reveal how the pressure equilibration between the two scales influences fracture propagation and lag-zone evolution in double-porosity media.
Second, extending the fracture flow to multiphase conditions, such as accounting for a gas phase in the lag zone or an immiscible displacing fluid, would broaden the applicability of the framework to scenarios such as supercritical CO$_2$ injection and magma-driven fracturing.

\section*{Acknowledgments}

This research was funded in whole or in part by the Austrian Science Fund (FWF) 10.55776/PIN9246524. The first author (TY) would like to thank Dr. Weihong Yuan for identifying the anisotropic feature of the Biot coefficient tensor.
		
\section*{Conflict of interest}
The authors declare that they have no known competing financial interests or personal relationships that could have appeared to influence the work reported in this paper.

\section*{Appendices}

\begin{appendices}
\numberwithin{equation}{section}
\section{Derivation of the Helmholtz free energy $\psi$ for the fluid-filled dual continuum} \label{ap:free energy}

Considering randomly distributed mesocracks (Fig.~\ref{fig:scale-seperation}) and fully saturated medium, the potential energy of the two-scale double-porosity material derived by \cite{pichler2010cracking} simplifies to
\begin{equation}
    \psi^\ast (\boldsymbol{\varepsilon}, p_\mathrm{c}, p_\mathrm{p}) = \frac{1}{2} \left( \boldsymbol{\varepsilon}:\mathbb{C}_\mathrm{eff}:\boldsymbol{\varepsilon}  -\frac{p_\mathrm{c}^2}{N_\mathrm{cc}} -\frac{p_\mathrm{p}^2}{N_\mathrm{pp}}\right) - p_\mathrm{c} \boldsymbol{\alpha}_\mathrm{c}:\boldsymbol{\varepsilon} - p_\mathrm{p}\boldsymbol{\alpha}_\mathrm{p}:\boldsymbol{\varepsilon} -\frac{p_\mathrm{c} p_\mathrm{p}}{N_\mathrm{pc}}
    .
    \label{eq:potential energy}
\end{equation}

According to the Legendre transform, the free energy functional $\psi$ as a function of the control variables $\boldsymbol{\varepsilon}$, $\zeta_\mathrm{c}$ and $\zeta_\mathrm{p}$ can be written as
\begin{equation}
    \psi (\boldsymbol{\varepsilon}, \zeta_\mathrm{c}, \zeta_\mathrm{p}) = \psi^\ast + p_\mathrm{c}\zeta_\mathrm{c} + p_\mathrm{p} \zeta_\mathrm{p}
    \label{eq:free energy}
\end{equation}
Substituting Eqs.~\eqref{eq:zeta_c}, \eqref{eq:zeta_p}, and \eqref{eq:potential energy} into Eq.~\eqref{eq:free energy}, we obtain Eq.~\eqref{eq:psi}.

\section{Simplification of the fixed-stress stabilization coefficients} \label{ap:fixed stress}

This appendix presents the detailed derivation leading from Eq.~\eqref{eq:fixed stress 2} to Eq.~\eqref{eq:fixed stress 3}.
The goal is to express the operator $\mathbb{S}_\mathrm{eff}: \boldsymbol{\alpha}_\mathrm{c} :\mathbb{C}_\mathrm{eff}$ appearing in Eq.~\eqref{eq:fixed stress 2} and $\mathbb{S}_\mathrm{eff}: \boldsymbol{\alpha}_\mathrm{p} :\mathbb{C}_\mathrm{eff}$ in Eq.~\eqref{eq:fixed stress p} solely in terms of Biot coefficient tensors defined in Eqs.~\eqref{eq:alpha_c_final}--\eqref{eq:alpha_p_final}.

We first establish that 
\begin{equation}
    \mathbb{S}_\mathrm{eff}:(\boldsymbol{\delta}:\mathbb{C}_\mathrm{eff}) = \boldsymbol{\delta} .\label{eq:ap_eq1}
\end{equation}
In index notation, the left-hand side reads
\begin{equation*}
    \left[\mathbb{S}_\mathrm{eff}:(\boldsymbol{\delta}:\mathbb{C}_\mathrm{eff})\right]_{ij}
    = (S_\mathrm{eff})_{ijkl}\,\delta_{mn}(C_\mathrm{eff})_{mnkl}.
\end{equation*}
Invoking the major symmetry of $\mathbb{C}_\mathrm{eff}$, i.e., $(C_\mathrm{eff})_{mnkl} = (C_\mathrm{eff})_{klmn}$, and the inversion relation $\mathbb{S}_\mathrm{eff}:\mathbb{C}_\mathrm{eff} = \mathbb{I}^\mathrm{sym}$, i.e., $(S_\mathrm{eff})_{ijkl}(C_\mathrm{eff})_{klmn} = (I^\mathrm{sym})_{ijmn}$, gives
\begin{equation*}
    \left[\mathbb{S}_\mathrm{eff}:(\boldsymbol{\delta}:\mathbb{C}_\mathrm{eff})\right]_{ij}
    = (I^\mathrm{sym})_{ijmn}\,\delta_{mn}
    = \tfrac{1}{2}(\delta_{im}\delta_{jn}+\delta_{in}\delta_{jm})\delta_{mn}
    = \delta_{ij},
\end{equation*}
hence $\mathbb{S}_\mathrm{eff}:(\boldsymbol{\delta}:\mathbb{C}_\mathrm{eff}) = \boldsymbol{\delta}$.

Applying $\mathbb{S}_\mathrm{eff}:(\cdot)$ to both sides of Eq.~\eqref{eq:alpha_c_final} yields
\begin{equation*}
    \mathbb{S}_\mathrm{eff}:\boldsymbol{\alpha}_\mathrm{c}
    = \mathbb{S}_\mathrm{eff}:\boldsymbol{\delta} - \frac{\mathbb{S}_\mathrm{eff}:(\boldsymbol{\delta}:\mathbb{C}_\mathrm{eff})}{3k_\mathrm{m}}
    = \mathbb{S}_\mathrm{eff}:\boldsymbol{\delta} - \frac{\boldsymbol{\delta}}{3k_\mathrm{m}},
\end{equation*}
Since the major symmetry of $\mathbb{S}_\mathrm{eff}$ implies $\mathbb{S}_\mathrm{eff}:\boldsymbol{\delta} = \boldsymbol{\delta}:\mathbb{S}_\mathrm{eff}$ as second-order tensors, the above relation is equivalently
\begin{equation}
   \mathbb{S}_\mathrm{eff}:\boldsymbol{\alpha}_\mathrm{c} =  \boldsymbol{\delta}:\mathbb{S}_\mathrm{eff} - \frac{\boldsymbol{\delta}}{3k_\mathrm{m}} = \boldsymbol{\alpha}_\mathrm{c} :\mathbb{S}_\mathrm{eff}.
    \label{eq:fs_identity}
\end{equation}

Substituting Eq.~\eqref{eq:ap_eq1} into the two contraction terms of Eq.~\eqref{eq:fixed stress 2} and Eq.~\eqref{eq:fixed stress p} gives
\begin{align*}
    \mathbb{S}_\mathrm{eff}: \boldsymbol{\alpha}_\mathrm{c} :\mathbb{C}_\mathrm{eff} &= \mathbb{S}_\mathrm{eff}: \boldsymbol{\delta} :\mathbb{C}_\mathrm{eff} - \frac{\mathbb{S}_\mathrm{eff}: \boldsymbol{\delta} :\mathbb{C}_\mathrm{eff}:\mathbb{C}_\mathrm{eff}}{3k_\mathrm{m}}
    = \boldsymbol{\delta} - \frac{ \boldsymbol{\delta} :\mathbb{C}_\mathrm{eff}}{3k_\mathrm{m}}
     = \boldsymbol{\alpha}_\mathrm{c}, \\
    \mathbb{S}_\mathrm{eff}: \boldsymbol{\alpha}_\mathrm{p} :\mathbb{C}_\mathrm{eff} &= \alpha_\mathrm{m} \frac{\mathbb{S}_\mathrm{eff}: \boldsymbol{\delta} :\mathbb{C}_\mathrm{eff}:\mathbb{C}_\mathrm{eff}}{3k_\mathrm{m}}
    = \alpha_\mathrm{m} \frac{ \boldsymbol{\delta} :\mathbb{C}_\mathrm{eff}}{3k_\mathrm{m}}
     = \boldsymbol{\alpha}_\mathrm{p}.
\end{align*}

Inserting these results into Eqs.~\eqref{eq:fixed stress 2} and \eqref{eq:fixed stress p} directly yields Eqs.~\eqref{eq:fixed stress 3} and \eqref{eq:fixed stress 4}.

\end{appendices}
	
	\bibliographystyle{apalike}
	\bibliography{reference} 
\end{document}